\documentstyle[prl,twocolumn,aps,epsf,eqsecnum]{revtex}
\begin{document}

 
\title{ Resonance contributions to HBT correlation radii }
\author{Urs Achim Wiedemann$^a$ and Ulrich Heinz$^{a,b}$}
 
\address{
   $^a$Institut f\"ur Theoretische Physik, Universit\"at Regensburg,\\
   D-93040 Regensburg, Germany \\
   $^b$CERN/TH, CH-1211 Geneva 23, Switzerland
}
 
\maketitle
 
\begin{abstract}
We study the effect of resonance decays on intensity interferometry 
for heavy ion collisions. Collective expansion of the source leads 
to a dependence of the two-particle correlation function on the pair 
momentum ${\bf K}$. This opens the possibility to reconstruct the 
dynamics of the source from the ${\bf K}$-dependence of the measured 
``HBT radii''. Here we address the question to what extent resonance 
decays can fake such a flow signal. Within a simple parametrization 
for the emission function we present a comprehensive analysis of the 
interplay of flow and resonance decays on the one- and two-particle 
spectra. We discuss in detail the non-Gaussian features of the 
correlation function introduced by long-lived resonances and the 
resulting problems in extracting meaningful HBT radii. We propose 
to define them in terms of the second order $q$-moments of the 
correlator $C({\bf q, K})$. We show that this yields a more reliable 
characterisation of the correlator in terms of its width and the 
correlation strength $\lambda$ than other commonly used fit procedures. 
The normalized fourth-order $q$-moments (kurtosis) provide a 
quantitative measure for the non-Gaussian features of the correlator.
At least for the class of models studied here, the kurtosis helps 
separating effects from expansion flow and resonance decays, and 
provides the cleanest signal to distinguish between scenarios with 
and without transverse flow.
\end{abstract} 
 

\section{Introduction}
\label{sec1}
 
The only known way to obtain direct experimental information on the 
space-time structure of the particle emitting source created in a
relativistic nuclear collision is through two-particle intensity 
interferometry \cite{GKW79,BGJ90}. This information is therefore 
indispensable for an assessment of theoretical models which try to 
reconstruct the final state of the collision from the measured single 
particle spectra and particle multiplicity densities in momentum space. 
Reliable estimates of the source {\em geometry} at particle freeze-out 
are crucial for an experimental proof that high energy heavy ion collisions 
can successfully generate large volumes of matter with high energy density. 
Direct information from two-particle correlations on the expansion 
{\em dynamics} at freeze-out further provides essential constraints
for theoretical models which extrapolate back in time towards the 
initial stages of the collision in order to make statements about a 
possible transition to deconfined quark matter.  
 
An important insight from recent theoretical research on 
Hanbury-Brown--Twiss (HBT) interferometry is that for dynamical sources 
which undergo collective expansion the HBT radius parameters, which
characterize the width of the two-particle correlation function, 
develop a dependence on the pair momentum 
\cite{P84,P86,MS88,B89,PCZ90,Marb,CL94,CSH95b,WSH96,HTWW96,WHTW96}. 
The detailed momentum dependence is somewhat model-dependent, and in 
general it is not simple \cite{WSH96}. Still, it opens the crucial 
possibility to extract dynamical information on the source from 
interferometry data. Unfortunately, the most abundant candidates for 
interferometry studies, charged pions, are strongly contaminated by 
decay products from unstable resonances some of which only decay long 
after hadron freeze-out \cite{GP89,Marb}. Such resonance decays were 
shown to introduce an additional momentum dependence of the HBT radius
parameters and of the intercept parameter \cite{Marb,CLZ96} which 
complicates the extraction of the expansion flow.  
 
A systematic approach towards extracting the expansion velocity from 
experimental HBT data thus presupposes a careful analysis of the 
interplay of flow and resonance decays on the gross features of the 
two-particle correlation function. This is the aim of the present 
paper. We will use for our analysis a simple analytical model for the 
source function, which assumes local thermalization at freeze-out and 
produces hadronic resonances by thermal excitation. The model 
incorporates longitudinal and transverse expansion as well as a finite 
duration of particle emission. The two most important parameters for
our considerations, the temperature and transverse expansion velocity 
at freeze-out, can be varied independently. Our study thus complements 
published HBT analyses of source functions generated by hydrodynamic 
simulations where freeze-out is implemented along a sharp hypersurface 
\cite{Marb} and which do not allow easily to gain physical intuition 
by a systematic variation of the model parameters. After freeze-out 
the resonances are allowed to decay according to an exponential proper 
time distribution along their trajectories, and the resulting emission 
functions of daughter particles (pions, kaons, etc.) are added to the 
direct emission function of particles of the same kind before 
calculating the correlation function. A discussion of the momentum 
dependence of resonance decay effects on the 1- and 2-particle spectra
requires the correct treatment of the decay phase-space 
\cite{Marb,H63,SKH91} and does not permit the simplifying 
approximations leading, e.g., to Eq.~(1) in Ref.~\cite{H96}.  
 
The paper is organized as follows: In Sec.~\ref{sec2} we review 
the extraction of space-time information on the source from Gaussian 
fits to the correlation function. This calculational scheme is then 
extended in Sec.~\ref{sec3} to include resonance decay contributions. 
The next three sections are devoted to a detailed model study based on
this formalism. In Sec.~\ref{sec4} we describe the model for the 
emission function including resonance contributions. Results for the 
one- and two-particle spectra are presented in Sec.~\ref{sec5},
and a general discussion of the effects from resonance decays on the 
shape of the correlation function is given there. In Sec.~\ref{sec6}
we then discuss in detail the practical difficulties posed by the 
non-Gaussian features in the correlation function due to long-lived 
resonances, by comparing different fitting procedures. This leads us 
in Sec.~\ref{sec7} to the alternative method of $q$-moments which 
provide a clean definition of the HBT radii and intercept parameter
even for non-Gaussian correlation functions. These HBT radii show much
weaker resonance decay effects than the ones obtained in Ref.~\cite{Marb}
by fitting a Gaussian function to a non-Gaussian correlator. The 
normalized fourth $q$-moment of the correlator, the kurtosis, provides
a quantitative measure for the deviations from a Gaussian shape as e.g.
induced by resonance decays. We will show that, at least within 
the general class of source models studied here, the simultaneous study
of the pair momentum dependence of the HBT radii, the intercept parameter
and the kurtosis allows for a relatively clean separation of flow and
resonance decay effects. We summarize our findings in Sec.~\ref{sec8}.
The Appendix contains some background for readers interested in the
technical details.
 
\section{Gaussian parametrizations of the correlation function}
\label{sec2}
 
For a given model for the emission function $S(x,p)$ 
and assuming incoherent particle production as well as plane wave 
propagation, the invariant momentum spectrum and two-particle HBT 
correlation function are given by \cite{S73,P84,CH94}
  \begin{eqnarray}
  \label{2.1}
    E_p {dN\over d^3p} &=& \int d^4x\, S(x,p) ,
  \\
  \label{2.2}
     C({\bf q},{\bf K}) &\approx& 1 + 
     {\left\vert \int d^4x\, S(x,K)\, e^{iq{\cdot}x}\right\vert^2 
      \over
      \left\vert \int d^4x\, S(x,K)\right\vert^2 }
    \nonumber \\
     &=& 1 + \left\vert \left\langle e^{i q{\cdot}x} 
           \right\rangle \right\vert^2 \, ,
   \\
   \label{2.3}
   \langle f(x) \rangle &\equiv& \langle f(x) \rangle (K) = 
   {\int d^4x\, f(x) \, S(x,K) \over \int d^4x \, S(x,K)} \, .
  \end{eqnarray}
Eq.~(\ref{2.2}) is written down for identical bosons, and $q = p_1 - 
p_2$, $K = \textstyle{1\over 2} (p_1 + p_2)$, with $p_1$, $p_2$ 
on-shell such that $K{\cdot}q =0$. 
$\langle f(x) \rangle \equiv \langle f(x) \rangle (K)$  denotes
the ($K$-dependent) average of an arbitrary space-time function 
with the emission function $S(x,K)$. As long as the emission function
is sufficiently Gaussian \cite{WSH96} one can approximate 
 \begin{equation}
 \label{2.4}
   C({\bf q},{\bf K}) \approx 1 + \exp\left[ - q_\mu q_\nu 
   \langle \tilde x^\mu \tilde x^\nu \rangle (K) \right]\, ,
 \end{equation}
where $\tilde x^\mu(K) = x^\mu - \langle x^\mu \rangle \equiv 
x^\mu - \bar x^\mu(K)$ is the distance to the point $\bar x(K)$ of 
maximum emissivity of particles with momentum $K$ in the source 
(the so-called ``saddle point" of the source for particles with 
momentum $K$). In this approximation the two-particle correlation 
function is completely determined by its Gaussian widths which in turn
are directly given by the ($K$-dependent) space-time variances 
$\langle \tilde x^\mu \tilde x^\nu \rangle$ of the emission function.
The latter define the size of regions of homogeneity in the source 
\cite{MS88,CSH95b,CSH95a,AS95} which effectively contribute to the
Bose-Einstein correlations. Finer space-time structures of the source 
show up in deviations of the correlator from a Gaussian shape.  
 
In previous studies of analytically given emission functions,
the correlator was sufficiently Gaussian to base all investigations 
on (\ref{2.4}). Then one proceeds as follows: Due to the on-shell 
constraint $K{\cdot}q=0$ only three of the four components of $q$ 
which appear in the exponent are independent. The dependent component must be 
eliminated using the relation
 \begin{equation}
 \label{2.5}
   q^0 = \bbox{\beta}\cdot {\bf q} = \beta_\perp q_o + \beta_l q_l\, .
 \end{equation}
Here $\bbox{\beta} = {\bf K}/K^0 \approx {\bf K}/E_K$, with
$E_K=\sqrt{m^2+{\bf K}^2}$, is approximately the velocity of the pair, 
and we used the convention that $l$ denotes the ``longitudinal" (beam)
direction ($z$-axis), $o$ denotes the orthogonal ``outward" direction
($x$-axis) which is oriented such that ${\bf K} = (K_\perp,,0,K_l)$ 
lies in the $x$-$z$-plane. Correspondingly $\bbox{\beta}$ has no 
$y$-component in the third Cartesian direction, the ``sideward" 
direction: $\beta_s=0$. Due to the mass-shell constraint (\ref{2.5}), 
the inverse of the Fourier transform in (\ref{2.2}) is not unique. 
The missing information required for the reconstruction of the 
(Gaussian) source in space-time from the measurable (Gaussian) HBT 
radii must thus be provided by model assumptions.
 
In this paper we will deal only with azimuthally symmetric sources
for which the correlation function is symmetric under $q_s \to 
-q_s$ \cite{CNH95}. Specifically, we will discuss two Gaussian
parametrizations of $C$:
 
{\bf 1.} The Cartesian parametrization \cite{CSH95a} is obtained by 
using (\ref{2.5}) to eliminate $q^0$ in (\ref{2.4}): 
  \begin{eqnarray}
       C({\bf q},{\bf K})
       &=& 1 + \lambda\,
            \exp\left[ - q_s^2\, R_s^2({\bf K}) - q_o^2\, R_o^2({\bf K})
           - q_l^2\, R_l^2({\bf K}) \right.
         \nonumber \\
         && \qquad \qquad \left.
           - 2 q_oq_l\, R_{ol}^2({\bf K})\right] \, .
  \label{2.6}
  \end{eqnarray}
The corresponding size parameters are given by the space-time 
variances \cite{HB95,CSH95a}:
 \begin{mathletters}
 \label{2.7}
 \begin{eqnarray}   
   R_s^2({\bf K}) &=& \langle \tilde{y}^2 \rangle \, ,
 \label{2.7a}\\
   R_o^2({\bf K}) &=& 
   \langle (\tilde{x} - \beta_\perp \tilde t)^2 \rangle \, ,
 \label{2.7b}\\
   R_l^2({\bf K}) &=& 
   \langle (\tilde{z} - \beta_l \tilde t)^2 \rangle \, ,
 \label{2.7c}\\
   R_{ol}^2({\bf K}) &=& 
   \langle (\tilde{x} - \beta_\perp \tilde t)
           (\tilde{z} - \beta_l \tilde t) \rangle \, .
 \label{2.7d} 
 \end{eqnarray}
 \end{mathletters}
For a detailed discussion of the meaning of these standard HBT 
parameters, in particular of the out-longitudinal ($ol$) 
cross term~\cite{CSH95a}, and how they mix spatial and temporal 
aspects of the source, see Refs.~\cite{CSH95b,WSH96,CNH95}.
 
{\bf 2}. If one eliminates in (\ref{2.4}) $q_o$ and $q_s$ in terms of 
$q_{\perp} = \sqrt{q_o^2 + q_s^2}$, $q^0$, and $q_l$ one arrives at
the Yano-Koonin-Podgoretski\u\i\ (YKP) parametrization 
\cite{YK78,P83,CNH95,HTWW96}
 \begin{eqnarray}
   C({\bf q},{\bf K}) &=&
       1 +  \lambda\,\exp\left[ - R_\perp^2({\bf K})\, q_{\perp}^2 
                       - R_\parallel^2({\bf K}) (q_l^2 - {q^0}^2 )
                       \right.
                       \nonumber \\
            && \qquad \left.
               - \left( R_0^2({\bf K}) + R_\parallel^2({\bf K})\right)
                 \left(q{\cdot}U({\bf K})\right)^2
                \right]  ,
 \label{2.8}
 \end{eqnarray}
where $U({\bf K})$ is a ($K$-dependent) 4-velocity with only a 
longitudinal spatial component:
 \begin{equation}
 \label{2.9}
   U({\bf K}) = \gamma({\bf K}) \left(1, 0, 0, v({\bf K}) \right) ,
   \ \ \text{with} \ \
   \gamma = {1\over \sqrt{1 - v^2}}\, .
 \end{equation}
The YKP parameters $R_\perp^2({\bf K})$, $R_0^2({\bf K})$, and 
$R_\parallel^2({\bf K})$ extracted from such a fit do not depend on 
the longitudinal velocity of the observer system in which the 
correlation function is measured. They can again be expressed in terms
of the space-time variances ${\langle{\tilde{x}_\mu\tilde{x}_\nu}\rangle}$
\cite{HTWW96}, and take their simplest form in the frame where $v({\bf K})$ 
vanishes \cite{CNH95,HTWW96,WHTW96} (the approximation in the last two 
expressions are discussed in \cite{CNH95,WHTW96}):
 \begin{mathletters}
 \label{2.10}
 \begin{eqnarray}   
 \FL
   R_\perp^2({\bf K}) &=& R_s^2({\bf K}) = \langle \tilde{y}^2 \rangle \, ,
 \label{2.10a} \\
   R_\parallel^2({\bf K}) &=& 
   \left\langle \left( \tilde z - (\beta_l/\beta_\perp) \tilde x
                \right)^2 \right \rangle   
     - (\beta_l/\beta_\perp)^2 \langle \tilde y^2 \rangle 
     \nonumber \\
     &\approx& \langle \tilde z^2 \rangle \, ,
 \label{2.10b} \\
   R_0^2({\bf K}) &=& 
   \left\langle \left( \tilde t - \tilde x /\beta_\perp
                \right)^2 \right \rangle 
    - \langle \tilde y^2 \rangle /\beta_\perp^2
    \approx \langle \tilde t^2 \rangle \, .
 \label{2.10c}
 \end{eqnarray}
 \end{mathletters}
\narrowtext
The expressions (\ref{2.7},\ref{2.10}) for the HBT parameters are useful
for two reasons: 
(i) They result in an appreciable technical simplification 
because instead of the Fourier transform (\ref{2.2}) only a small number 
of 4-dimensional real integrals over the source function must be evaluated
to completely determine the correlation function. Their accuracy has been 
checked in \cite{WSH96} for models of the type to be used below and, in 
the absence of resonance decays, for hydrodynamic sources with a sharp 
freeze-out hypersurface in \cite{Sch96}. 
(ii) They provide an intuitive understanding of which space-time features 
of the source are reflected by the various $q$-dependencies of the 
correlator. However, their range of validity is limited by 
the fact that strictly speaking the space-time variances determine 
only the curvature of the correlator at ${\bf q}=0$:
  \begin{equation}
    {\langle{ (\tilde{x}_i - \beta_i\tilde{t})
        (\tilde{x}_j - \beta_j\tilde{t})}\rangle}
    = - {1\over 2} 
        {\partial^2 C({\bf q},{\bf K})\over \partial q_i\, \partial q_j} 
    \Bigg\vert_{{\bf q} = 0}\, .
  \label{2.11}
  \end{equation}
This agrees with the widths of the correlator if and only if 
$C({\bf q},{\bf K})$ is Gaussian. We will see that resonance decays 
can lead to appreciable non-Gaussian features in the correlation 
function. If this is the case, Eqs.~(\ref{2.7}) and (\ref{2.10}) no 
longer provide quantitatively reliable expressions for the observed 
half widths of the correlator. A detailed discussion will follow in 
Sec.~\ref{sec5}.
 
\section{Resonance decay contributions}
\label{sec3}
 
We concentrate on charged pion ($\pi^+\pi^+$ or $\pi^-\pi^-$) 
correlations. In the presence of resonance decays, the emission 
function is the sum of a direct term plus one additional term for 
each resonance decay channel with a pion of the desired charge 
in the final state: 
  \begin{equation}
     S_{\pi}(x,p) = S_{\pi}^{\rm dir}(x,p) + 
     \sum_{r\ne\pi} S_{r\to \pi}(x,p)\, .
  \label{3.1}
  \end{equation}
Note that the sum is over decay channels, not just over resonances.
We compute the emission functions $S_{r\to \pi}(x,p)$ for the decay 
pions from the direct emission functions $S_r^{\rm dir}(X,P)$ for 
the resonances taking into account the correct decay kinematics for 2-
and 3-body decays,   
  \begin{equation}
  \label{3.2}
        S_{r\to\pi}(x,p) = \sum_\pm \int_{\bf R} 
        \int_0^{\infty}{d\tau}\, \Gamma e^{-\Gamma\tau} 
        S_r^{\rm dir} \left( x -{P^\pm\over M} \tau,P^\pm \right) \, .
 \end{equation}
From now on capital letters denote variables 
associated with the parent resonance, while lowercase letters denote 
pion variables. Here, $\Gamma$ is the total decay width of the
resonance, and $\sum_\pm \int_{\bf R}$ goes over the kinematically
allowed resonance momenta as described in Appendix~\ref{appa}. Please 
note that the momenta $p$ and $P^\pm$ in this expression are in 
general different, in contrast to the approximation used in Eq.~(1) of 
Ref.~\cite{H96}. This is important for the following discussion of the 
momentum dependence of the correlator.  
 
\begin{figure}[h]\epsfxsize=8cm 
\centerline{\epsfbox{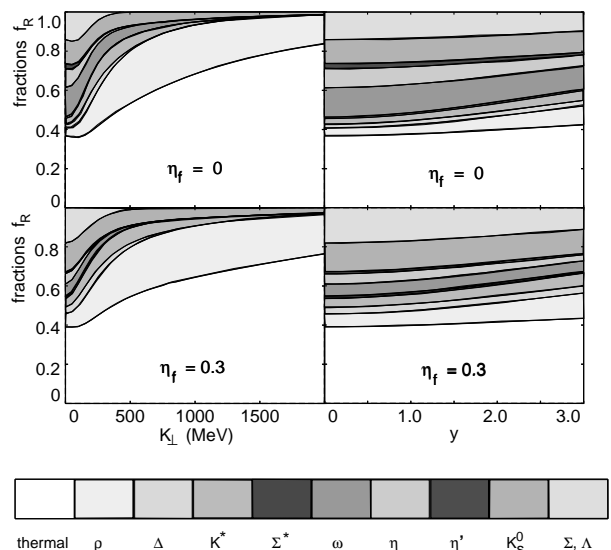}}
\caption{
The resonance fractions $f_r(y,p_\perp)$ according to 
Eq.~(\protect\ref{3.6}) for $T = 150$ MeV. Upper row: no transverse
flow, $\eta_f=0$; lower row: $\eta_f=0.3$. Left column: $f_r$ as
a function of transverse momentum at central rapidity; right column: 
$f_r$ as function of rapidity at $p_\perp=0$.
}\label{F1}
\end{figure}

The complete two-particle correlation function is then given by
  \begin{eqnarray}
    C({\bf q},{\bf K}) &=& 1+ 
    {\vert\tilde S_{\pi}^{\rm dir}(q,K)\vert^2
     \over 
     \vert \tilde S_\pi(0,K) \vert^2}
   \nonumber \\
     && + 2\,{\sum_{r\ne\pi} {\rm Re} [\tilde S_{\pi}^{\rm dir}(q,K)
                         \tilde S_{r\to\pi}(q,K) ]
     \over 
     \vert \tilde S_\pi(0,K) \vert^2}
   \nonumber \\
     && + {\vert \sum_{r\ne\pi} \tilde S_{r\to\pi}(q,K)\vert^2 
     \over 
     \vert \tilde S_\pi(0,K) \vert^2} \, ,
  \label{3.3}
  \end{eqnarray}

where the denominator includes all resonance contributions according 
to (\ref{3.1}). The last term in the numerator can be neglected
if resonance production is small \cite{G77}. However, in 
ultrarelativistic heavy ion collisions a major fraction of all 
final state pions stem from resonance decays (see Fig.~\ref{F1})
and this ``Grassberger approximation'' cannot be used. 
 
For later reference, we extend the expressions given 
in Sec.~\ref{sec2} for the HBT parameters in terms of space-time 
variances of the source to include resonance decay contributions:
  \begin{equation}
    \langle \tilde x_\mu \tilde x_\nu \rangle(K) = 
    {\sum_r \int d^4x\, \tilde x_\mu \tilde x_\nu \, S_{r\to\pi}(x,K) 
     \over
     \sum_r \int d^4x\, S_{r\to\pi}(x,K)} \, .
    \label{3.4}
  \end{equation}
Here the sum runs over all contributions, including the direct 
pions. It is instructive to rewrite the average over the emission 
function in the following form:
  \begin{eqnarray}
  \label{3.5}
    \langle x_\nu \rangle (K)  
    &=& \sum_r f_r(K) \ 
        \langle x_\nu \rangle_r (K)\, ,
  \nonumber\\
    \langle x_\mu x_\nu \rangle (K)  
    &=& \sum_r f_r(K) \ 
        \langle x_\mu x_\nu \rangle_r (K)\, ,
  \end{eqnarray}
where we introduced the single-particle fractions \cite{Marb}
  \begin{eqnarray}
    f_r(K) &=& {\int d^4x\, S_{r\to\pi}(x,K) \over
                     \sum_r \int d^4x\, S_{r\to\pi}(x,K)} 
                   = {dN_\pi^r/d^3K \over dN_\pi^{\rm tot}/d^3K}\, ;
    \nonumber \\
    && \sum_r f_r(K) = 1\, .
  \label{3.6}
  \end{eqnarray}
These give the fraction of single pions with momentum ${\bf K}$ 
resulting from decay channel $r$. We also defined the average $\langle 
\dots \rangle_r$ with the effective pion emission function arising from
this particular channel: 
  \begin{equation}
  \label{3.7}
    \langle \dots \rangle_r(K) = 
    {\int d^4x \dots S_{r\to\pi}(x,K) \over
     \int d^4x\, S_{r\to\pi}(x,K)} \, .
  \end{equation}
The variances (\ref{3.4}) can then be rewritten as
  \begin{equation}
    \langle \tilde x_\mu \tilde x_\nu \rangle 
    = \sum_r f_r\, \langle \tilde x_\mu \tilde x_\nu \rangle_r 
    + \sum_{r,r'} f_r (\delta_{r,r'} - f_{r'}) 
    \langle x_\mu \rangle_r \langle x_\nu \rangle_{r'}\, . 
  \label{3.8}
  \end{equation}
The first term has a simple intuitive interpretation: each resonance 
decay channel $r$ contributes an effective emission function 
$S_{r\to\pi}$. The full variance is calculated by weighting the 
variance (homogeneity length) of the emission function from a 
particular decay channel $r$ with the fraction $f_r$ with which this 
channel contributes to the single particle spectrum. However,
the different effective emission functions $S_{r\to\pi}(x,p)$ 
have in general different saddle points; this gives rise to the second 
term in (\ref{3.8}) which somewhat spoils its intuitive interpretation.
 
Also, the full emission function (\ref{3.1}) is a superposition of 
sources with widely differing sizes since long-lived resonances contribute 
long exponential tails to the emission function $S_{r\to\pi}$ 
\cite{Marb,CLZ96}. It is easy to see that this leads to non-Gaussian
correlation functions: Consider a simple 1-dimensional toy model where 
the emission function is a sum of two Gaussian terms, one of
width $R_{\rm dir}$ for direct pions and one of width
$R_{\rm halo}$ for pions from a resonance, with weights $1-\epsilon$ 
and $\epsilon$, respectively:
 \begin{eqnarray}
   S_\pi(x,K) &=& S_\pi^{\rm dir}(x,K) + S_{r\to\pi}(x,K) 
   \nonumber \\
   &=& (1-\epsilon)\, e^{-x^2/(2R_{\rm dir}^2)} + 
   \epsilon\, e^{-x^2/(2R_{\rm halo}^2)}\, .
 \label{3.9}
 \end{eqnarray}
According to (\ref{3.3}) the correlator is then a superposition of three
Gaussians which for $R_{\rm halo} \gg R_{\rm dir}$ have
very different widths:
 \begin{eqnarray}
   C(q,K) - 1 &=& (1-\epsilon)^2\, e^{-R_{\rm dir}^2 q^2} + 
   \epsilon^2\, e^{-R_{\rm halo}^2 q^2} 
   \nonumber \\
   && +  2 \epsilon (1-\epsilon) \,
   e^{-(R_{\rm dir}^2 + R_{\rm halo}^2) q^2/2}\, .
 \label{3.10}
 \end{eqnarray}
Obviously, if $\epsilon$ is small, the rough structure of the 
correlator will be determined by the large and broad direct 
contribution. The two other contributions will, however, modify its 
functional form as follows: 
\newline
(i) If the resonance is shortlived such that $R_{\rm halo} \approx
R_{\rm dir}$, its effect on the correlator will be minor; its shape 
will remain roughly Gaussian, with a width somewhere between $1/R_{\rm 
dir}$ and $1/R_{\rm halo}$, depending on the weight $\epsilon$ of 
the resonance contribution.  
\newline
(ii) If the resonance lifetime and thus $R_{\rm halo}$ are extremely 
large, the second and third term in (\ref{3.10}) will be very narrow 
and, due to the finite two-track resolution of every experiment, may 
escape detection; then the correlator looks again Gaussian with a 
width $1/R_{\rm dir}$, but at $q=0$ it will not approach the value 2, 
but $1 + (1-\epsilon)^2 < 2$. The correlation appears to be 
incomplete, with a ``correlation strength" $\lambda = (1-f_r)^2 
= (1-\epsilon)^2$. 
\newline
(iii) If the resonance lifetime is in between such 
that $R_{\rm halo} \gg R_{\rm dir}$ but $1/R_{\rm halo}$ is still 
large enough to be experimentally resolved, all three Gaussians 
contribute, and the full correlator deviates strongly from a single 
Gaussian.  
%
%
In cases (ii) and (iii) the space-time variances calculated from 
Eq.~(\ref{3.8}) yield misleading or outright wrong results for the 
width of the correlation function. They give the curvature of the 
correlator at ${\bf q}=0$: 
 \begin{eqnarray}
   \langle \tilde x^2 \rangle &=& 
   f_{\rm dir} \langle \tilde x^2 \rangle_{\rm dir} + 
   f_{\rm halo} \langle \tilde x^2 \rangle_{\rm halo} = 
   (1-\epsilon) R_{\rm dir}^2 + \epsilon R_{\rm halo}^2 
   \nonumber \\
   &=& -{1\over 2}
   \left. {\partial^2 C(q)\over \partial q^2} \right\vert_{q=0} \, .
 \label{3.11}
 \end{eqnarray}
In case (ii), for not too small values of $\epsilon$, this is dominated 
by the second term although the resonance contribution is not even 
visible in the measured correlator! On a quantitative level, the 
situation is not very much better for case (iii) (see 
Sec.~\ref{sec6a1} for a more detailed discussion).  
 
However, if the resonances can be clearly separated into two distinct 
classes, one with very short lifetimes of order 1 fm/$c$, the other 
with very long lifetimes of order 100 fm/$c$ or longer, then 
space-time variances can again provide an accurate measure for the 
width of the correlation function. To achieve this, one must leave out 
the long-lived resonances from the sum over $r$ in (\ref{3.8}), i.e. 
one restricts the calculation of the space-time variances to the 
``core'' of the emission function from direct pions and short-lived 
resonances \cite{He96,Cs96}. Since the contribution from long-lived 
resonances to the correlator cannot be resolved experimentally (while 
they do contribute to the single-particle spectra), one includes them 
via a reduced correlation strength $\lambda$: 
 \begin{equation}
 \label{3.12}
   \lambda(K) = \left( 
   1 - \sum_{r = {\rm longlived}} f_r(K) \right)^2\, .
 \end{equation}
The $K$-dependence of $\lambda$ will be discussed in Sec.~\ref{sec5}.
 
The real problem comes from resonances with an intermediate lifetime. 
They cause appreciable deviations from a Gaussian behaviour for the 
correlator and cannot be reliably treated by the method of space-time 
variances. In nature there is only one such resonance, the $\omega$ meson
with its 23.4 fm/$c$ lifetime. At low $K_\perp$ it contributes 
up to 10\% of all pions ($f_\omega({\bf K}=0) \approx 0.1$), and their
non-Gaussian effects on the correlator can be clearly seen. 
They will be discussed extensively in Sections~\ref{sec6},\ref{sec7}.
 
\section{A simple model for the emission function}
\label{sec4}
 
As discussed after Eq.~(\ref{2.5}), a completely model-independent 
HBT analysis is not possible. In this Section we define a simple 
model for the emission function in relativistic nuclear collisions 
which will be used in the rest of the paper for quantitative studies.
It has been used extensively in the literature 
\cite{CL94,CSH95b,WSH96,HTWW96,WHTW96,CNH95}, and we present a simple 
extension to include resonance production. It implements the essential 
features expected from sources created in nuclear collisions: It 
assumes local thermalization prior to freeze-out and incorporates 
its collective expansion in the longitudinal and transverse directions. 
On the geometric side, the source has a finite size in the spatial and 
temporal directions, i.e. it implements a finite, but non-zero duration 
for particle emission. 
 
The emission function for particle species $r$ is taken as
  \begin{eqnarray}
    S^{\rm dir}_r(x,P) &=& {2J_r + 1 \over (2\pi)^3}\,
   M_\perp \cosh(Y-\eta) 
   \nonumber \\
   && \times 
   \exp\left(- {P \cdot u(x) - \mu_r \over T} \right)\,  H(x)
       \label{4.1}
  \end{eqnarray}
where 
 \begin{equation}
 \label{4.2}
   H(x) = {1\over \pi (\Delta\tau)}\,
          \exp\left( - {r^2\over 2 R^2} 
                     - {(\eta-\eta_0)^2\over 2 (\Delta\eta)^2}
                     - {(\tau-\tau_0)^2 \over 2 (\Delta\tau)^2}
                 \right) \, , 
 \end{equation}
with proper time $\tau = \sqrt{t^2 - z^2}$ and space-time rapidity
$\eta = \textstyle{1\over 2} \log{\lbrack{ (t+z)/(t-z) }\rbrack}$.
The physical meaning of the parameters has been explained in detail in 
Refs.~\cite{CL94,CSH95b,WSH96,HTWW96,WHTW96,CNH95} to which we refer
the reader. The only new ingredients are a factor $2J_r+1$ for
the spin degeneracy (due to charge identification 
in the experiment each isospin state must be treated separately),
and a chemical potential $\mu_r$ for each resonance $r$. This means
that all particles are assumed to freeze out with the same geometric 
characteristics and the same collective flow, superimposed by thermal 
motion with the same temperature. The possible consequences of 
particle-specific freeze-out \cite{N82,HLR87} will have to be 
discussed elsewhere.  
 
For later reference we note that the function $H(x)$ is normalized to 
the total comoving 3-volume according to
  \begin{eqnarray}
  \label{4.3}
    \int d^4x\, H(x) &=& \pi r^2_{\rm rms} \cdot 2 \tau_0 \eta_{\rm rms}\, ,
  \\
  \label{4.4}
    r^2_{\rm rms} = 2 R^2 = x^2_{\rm rms} + y^2_{\rm rms}\, ,
    && \qquad
    \eta_{\rm rms} = \Delta\eta\, .
 \end{eqnarray}
Note that the rms widths in $x$- and 
$y$-direction are each given by $R$. If the Gaussians in $H(x)$ were
replaced by box functions \cite{SSH93,SH92}, the equivalent box 
dimensions (with the same rms radii) would be $\tilde R=2R$, $\tilde 
\eta = \sqrt{3}\, \Delta \eta$.  
 
For the flow profile we assume \cite{HTWW96} Bjorken scaling in the 
longitudinal direction, $v_l=z/t$, and a linear transverse flow 
rapidity profile \cite{SH93}:
 \begin{equation}
 \label{4.5}
   \eta_t(r) = \eta_f {r\over R}\, .
 \end{equation}
In spite of the longitudinal boost-invariance of the flow, 
the source as a whole is not boost-invariant due to the finite 
extension in $\eta$ provided by the second Gaussian in (\ref{4.1}).  
 
Inserting the parametrization (\ref{A5}) for $P$ the emission 
function (\ref{4.1}) becomes \cite{WHTW96}
  \begin{eqnarray}
     S_r^{\rm dir}(x,P) 
     &=& { 2 J_r + 1 \over (2\pi)^3}\,
          M_T \cosh(Y-\eta)\, e^{\mu_r/T}\,H(x)
  \nonumber \\
     && \times \exp\left(-{M_T\over T}\cosh(Y-\eta)\cosh\eta_t(r)
       \right.
       \nonumber \\
       && \qquad \left.
            +{P_T\over T} \sinh\eta_t(r) \cos(\phi-\Phi)\right)\, .
  \label{4.6}
  \end{eqnarray}
The direct pion component $S_\pi^{\rm dir}(x,p)$ is obtained from 
this expression by setting $r=\pi$, $P=p$, $J_\pi=0$, $\mu_\pi=0$, 
and $\Phi=0$ (see (\ref{A4})).
This last condition reflects a choice for the orientation of the 
coordinate system such that the transverse momentum ${\bf p}_\perp$ 
of the decay pion lies in the $x-z$ plane. For the transverse 
momentum ${\bf P}_\perp$ of resonances which contribute pions with 
the same ${\bf p}_\perp$ as the directly emitted ones,  
in general a non-vanishing azimuthal angle $\Phi$ is required,
see Appendix~\ref{appa}.
 
The chemical potentials $\mu_r$ will be fixed by the assumption of 
chemical equilibrium at freeze-out. Then baryon number and strangeness 
conservation in the fireball demand the existence of two independent 
chemical potentials $\mu_B$ and $\mu_S$, with
\begin{table}
\begin{tabular}{ccccc}
Decay channel $r$ & $M$ (MeV) & $\Gamma$ (MeV) & $J$ & 
$b_{r\to\pi^-}$ \\
\hline
$\rho^- \to \pi^-  \pi^0$ & 770 & 150 & 1 & 1.0 \\
$\rho^0 \to \pi^-  \pi^+$ & 770 & 150 & 1 & 1.0 \\
& & & & \\
$\Delta^- \to \pi^-  n$ & 1232 & 115 & 3/2 & 1.0 \\                  
$\Delta^0 \to \pi^-  p$ & 1232 & 115 & 3/2 & $(1/3)\times 1.0$ \\ 
$\bar{\Delta}^{+} \to \pi^-  \bar{n}$ & 1232 & 115 & 3/2 
                  & $(1/3)\times 1.0$ \\
$\bar{\Delta}^{++} \to \pi^-  \bar{p}$& 1232 & 115 & 3/2 & 1.0 \\
& & & & \\
${K^*}^0 \to \pi^-  K^+$ & 892 & 50 & 1 & $(2/3)\times 1.0$ \\
${K^*}^- \to \pi^-  K^0$ & 892 & 50 & 1 & $(2/3)\times 1.0$ \\
& & & & \\
${\Sigma^*}^- \to \pi^-  \Lambda(1116)$ & 1385 & 36 & 1/2 & 0.88 \\
${\Sigma^*}^- \to \pi^-  \Sigma^0(1193)$ & 1385 & 36 & 1/2 & 
$(1/2)\times 0.12$ \\
${\Sigma^*}^0 \to \pi^-  \Sigma^+(1193)$ & 1385 & 36 & 1/2 & 
$(1/2)\times 0.12$ \\
$\bar{\Sigma}^*{}^+ \to \pi^-  \bar{\Lambda}(1116)$ & 1385 & 
36 & 1/2 & 0.88 \\
$\bar{\Sigma}^*{}^+ \to \pi^-  \bar{\Sigma}^0(1193)$ & 1385 & 36 & 1/2 &
$(1/2)\times 0.12$ \\
$\bar{\Sigma}^*{}^0 \to \pi^-  \bar{\Sigma}^-(1193)$ & 1385 & 36 & 1/2 &
$(1/2)\times 0.12$ \\
& & & & \\
$\omega \to \pi^-  \pi^+  \pi^0$ & 782 & 8.43 & 1 & 0.89 \\
& & & & \\
$\eta \to \pi^-  \pi^+  \pi^0$ & 547 & $1.2 \times 10^{-3}$ & 0 & 0.24 \\ 
& & & & \\
$\eta' \to \pi^+  \pi^-  \eta $ & 958 & 0.2 & 0 & 0.44 \\
& & & & \\
$K_S^0 \to \pi^+ \pi^-$ & 498 & $\approx 0$ & 0 &
$0.69$ \\
& & & & \\
$\Sigma^- \to \pi^- n$ & 1193 & $\approx 0$ & 1/2 &
$1.0$ \\
$\bar{\Sigma}^+ \to \pi^- \bar{n}$ & 1193 & $\approx 0$ & 1/2 &
$1.0$ \\
${\Sigma}^0 \to \gamma\Lambda \to p \pi^-$ & 1193 & $\approx 0$ & 1/2 &
$0.65$ \\
$\Lambda \to p \pi^-$ & 1116 & $\approx 0$  & 1/2 & $0.65$ \\
\end{tabular}
\caption{The resonance decay contributions to $\pi^-$ production 
considered in the present work. Where applicable the factor in front 
of the branching ratio is the Clebsch-Gordon coefficient for the 
particular decay channel.}
\label{T1}
\end{table}
 \begin{equation}   
 \label{4.7}
   \mu_r = b_r \mu_B + s_r \mu_S\, ,
 \end{equation}
where $b_r$ and $s_r$ are the baryon number and strangeness of 
resonance $r$, respectively. The condition of overall strangeness 
neutrality of the fireball allows to eliminate $\mu_S$ in terms of $T$ 
and $\mu_B$ \cite{LTHSR95}.

Unless stated otherwise, the numerical calculations~\cite{WH97b}
below are done with
the set of source parameters $T = 150$ MeV, $R = 5$ fm, $\Delta\eta = 1.2$,
$\tau_0 = 5$ fm/c, $\Delta\tau = 1$ fm/c and $\mu_B = \mu_S = 0$. 
We will work in the fireball c.m. system and thus set $\eta_0=0$.

The resonance channels included are listed in Table~\ref{T1}. The
$\Sigma$(1193) and $\Lambda$(1116) are treated as one baryonic
resonance $Y$(1150) at an average mass of 1150 MeV. For simplicity 
the decay cascade $\Sigma^0 \to \gamma\Lambda \to p\pi^-$ is replaced by 
an effective two-particle decay $\Sigma^0 \to p\pi^-$, since the photon 
in the $\Sigma^0$-decay is known not to change the shape of the hyperon 
spectrum \cite{SSH93}. The $\pi^-$ decay contributions from the cascades
$\eta' \to ...$ $ + \eta \to \pi^- + ...$ and $\Sigma^* \to ... +$ $Y(1150) 
\to \pi^- + ...$ are taken into account by enhanced branching ratios 
for the $Y$ and $\eta$ decay channels. These crude approximations are 
not problematic because they concern quantitatively small contributions. 
The cascade decays just mentioned affect the intercept parameter on the 
level of a few percent; the ${\bf K}$- dependence of the HBT radius 
parameters remains essentially unaffected. $K_L^0$ decays are neglected 
because the long $K_L^0$ lifetime ($c\tau$ = 15.5 m) makes them invisible 
for most detectors.

\section{Results for one- and two-particle spectra}
\label{sec5}
 
We now present a quantitative analysis of the one- and two-particle 
spectra for the model described in Sec.~\ref{sec4}. 
Both types of spectra can be expressed in terms 
of the four-dimensional Fourier transforms of the direct emission 
functions, $\tilde S_r^{\rm dir}(q,P^{\pm})$, see Appendix~\ref{appa}. 
We show in 
Appendix~\ref{appb} how the latter can be reduced analytically 
to 2-dimensional integrals over $r$ and $\eta$:
  \begin{eqnarray}
    \tilde S_r^{\rm dir} (q,P^\pm) 
    &=& {(2J_r+1) \over \pi (2\pi)^{3/2} }\, 
        M_\perp \, \tau_0\, e^{{\mu_r\over T}}
        \int d\eta\, \left( 1 + i A_q \right) 
        \nonumber \\
        && \times \cosh(\eta-Y) \, 
        e^{i A \tau_0}\, e^{- {1\over 2} A^2 (\Delta\tau)^2} 
        e^{- {\eta^2 \over 2 (\Delta\eta)^2}}\,
  \nonumber \\
    && \times  \int_0^{\infty} r\, dr\, e^{-{r^2\over 2R^2}}\, 
             e^{-{M_\perp \over T}\cosh(\eta-Y)\cosh\eta_t} 
             \nonumber \\
             && \times
             I_0 \left( \sqrt{C-iD^\pm} \right)\, , 
  \label{5.1}
  \end{eqnarray}
where
  \begin{mathletters}
  \label{5.2}
  \begin{eqnarray}
     C(r) &=& {P_\perp^2 \over T^2} \sinh^2\eta_t(r) - r^2\,q_\perp^2
     \, ,
  \label{5.2a} \\
     D^\pm(r) &=& - 2\, r \, {P_\perp \over T} \sinh\eta_t(r)
         \nonumber \\
         && \times 
        \left( q_o \cos\Phi_\pm + q_s \sin\Phi_\pm \right)\, ,
  \label{5.2b} \\
     A(\eta) &=& \left( q^0 \cosh\eta - q_l \sinh\eta \right)\, ,
  \label{5.2c} \\
     A_q(\eta) &=& A(\eta) {(\Delta\tau)^2\over \tau_0}\, .
  \label{5.2d}
  \end{eqnarray}
  \end{mathletters}
The Bessel function $I_0$ arises from the $\phi$-integration while the 
terms containing $A$ and $A_q$ stem from the $\tau$-integration. 
Please note that the azimuthal rotation of the resonance transverse 
momentum ${\bf P}_\perp$ relative to the pion transverse momentum 
${\bf p}_\perp$ (which defines the $x$-axis of our coordinate system) 
enters only through the combination in brackets in Eq.~(\ref{5.2b});
the latter stems from the scalar product $q\cdot P^\pm$, see 
Eq.~(\ref{A19}). This means that the dependence on $\Phi_\pm$ can 
be shifted from ${\bf P}^\pm_\perp$ to ${\bf q}_\perp$ by a common 
rotation by the angle $\Phi_\pm$:
  \begin{eqnarray}
    && \tilde S_r^{\rm dir} \left(q^0,q_o,q_s,q_l;
       E_{_P},P_\perp\cos\Phi_\pm, 
       P_\perp\sin\Phi_\pm,P_{_L}\right) =
  \nonumber\\
    && \qquad
       \tilde S_r^{\rm dir} \left(q^0,
       q_o \cos\Phi_\pm + q_s \sin\Phi_\pm ,
       q_s \cos\Phi_\pm - \right.
     \nonumber \\
     && \qquad \, \, \left.
       q_o \sin\Phi_\pm ,
       q_l; E_{_P},P_\perp,0,P_{_L}\right) \, .
  \label{5.3}
  \end{eqnarray}
Note that this identity does not depend on the model for the emission 
function. It shows that the resonance decay kinematics leads to a mixing 
of the sideward and outward $q$-dependencies of the correlation 
functions that would be obtained from the resonances if one could use 
them for interferometry directly. This feature is lost in the 
approximation leading to Eq.~(2) in Ref.~\cite{H96}.
 
For the direct pion contribution, $\Phi_\pm$ is to be set 
to zero in (\ref{5.1}), (\ref{5.3}).  
 
\subsection{The resonance fractions $f_r(K)$}
\label{sec5a} 
 
The single particle momentum spectrum (\ref{2.1}) is the space-time 
integral over the emission function (\ref{3.1}),
  \begin{equation}
    {dN_\pi\over \pi\,dy\,dm_\perp^2} = 
    \int d^4x\, S_\pi(x,p) = \tilde S_\pi(q=0;m_\perp,y).
    \label{5.4}
  \end{equation}
It is thus given by the Fourier transform (\ref{5.1}) of the emission 
function at zero relative momentum. From this expression it is 
straightforward to evaluate the resonance fractions $f_r(y,m_\perp)$ 
of Eq.~(\ref{3.6}). For later referene they are shown in 
Fig.~\ref{F1}. At central rapidity and small transverse momentum in 
our model only about 40\% of the pions are emitted directly while more 
than half of the pions stem from resonance decays. The direct fraction 
increases rapidly with increasing transverse momentum, but very slowly 
with increasing longitudinal momentum resp. rapidity. In fact, most 
resonance fractions are nearly independent of rapidity \cite{Marb}. At 
large $p_\perp$ the resonance contributions to the single particle 
spectrum die out \cite{SKH91}. The largest resonance contribution 
comes from the $\rho$ meson, due to its relatively small mass and 
large spin degeneray factor. The $\eta$, which is still lighter, has 
no spin and a small branching ratio into pions. As can be seen in the 
lower row of Fig.~\ref{F1} the resonance fractions are only weakly 
affected by transverse flow: at small $p_\perp$ the direct fraction 
increases slightly while at large $p_\perp$ the tendency is opposite 
(see Sec.~\ref{sec5b}).  
 
\subsection{Single particle transverse momentum spectra}
\label{sec5b} 
 
Integrating (\ref{5.4}) over rapidity we obtain the single particle 
transverse momentum distribution:
  \begin{eqnarray}
    {dN_\pi\over dm_\perp^2}  
    &=& \pi \int dy\, \tilde S_\pi^{\rm dir}(0;y,m_\perp) 
    \nonumber \\
    && + \sum_{r\ne\pi} \pi \int dy\, \tilde S_{r\to \pi}(0;y,m_\perp) \, .
  \label{5.5}
  \end{eqnarray}
The resonance decay contributions are given according to Eqs.~(\ref{A18}) 
and (\ref{A19}) by
  \begin{equation}
  \label{5.6}
     \tilde S_{r\to\pi}(0;y,m_\perp) 
     = 2 M \int_{\bf R} S_r^{\rm dir}(0;Y,M_\perp) \, .
  \end{equation}
The factor 2 results from the sum over $\Phi_\pm$, noting that at $q=0$ 
the integrand is independent of $\Phi_\pm$ (see Appendix~\ref{appb}). 
Writing $Y=y + v\Delta Y$ (see Eq.~(\ref{A16})) where $\Delta Y$ is 
independent of $y$, the $y$-integration can be pulled through the 
integrals $\int_{\bf R}$ over the decay phase space, yielding~\cite{SKH91}
  \begin{equation}
  \label{5.7}
    {dN_\pi\over dm_\perp^2} = {dN_\pi^{\rm dir}\over dm_\perp^2}
     + \sum_{r\ne\pi} 2 M_r \int_{\bf R} {dN_r^{\rm dir} \over dM_\perp^2}\, .
  \end{equation}
The transverse momentum spectra of the directly emitted resonances $r$ 
are given by expression (\ref{B4a}) \cite{SKH91,SSH93}: 
  \begin{eqnarray}
  \label{5.8}
    {dN_r^{\rm dir}\over dM_\perp^2} 
    &=& {2J_r+1 \over 4\pi^2} \, 
        (2 \pi R^2 \cdot 2 \tau_0 \Delta\eta)\,
        e^{\mu_r/T}
  \nonumber\\     
    &&\times \, M_\perp 
        \int_0^\infty d\left({\xi^2\over 2}\right) e^{-\xi^2/2}\,
        K_1\left( {\textstyle{M_\perp\over T}}\cosh\eta_t(\xi) \right)
  \nonumber \\
    && \times \,
    I_0\left( {\textstyle{P_\perp\over T}}\sinh\eta_t(\xi) \right) \, ,
  \end{eqnarray}
where we substituted $\xi = r/R$ under the integral. Note that the 
geometric parameters $R$, $\Delta\eta$, $\tau_0$ of the source enter 
only in the normalization of the spectrum through the effective 
volume (\ref{4.3}). Thus the shape of the ($y$-integrated!) 
single-particle transverse momentum spectrum contains no information 
on the source geometry, in agreement with general arguments presented 
e.g. in \cite{HTWW96}. According to (\ref{5.7},\ref{5.8}), the unnormalized  
transverse momentum dependence is fully determined by the rest mass 
$M$, the temperature $T$ (or $T(\xi)$ if $T$ were $r$-dependent), and 
the transverse flow profile $\eta_t(\xi)=\eta_f \xi^n$.
 
\begin{figure}[h]\epsfxsize=7cm 
\centerline{\epsfbox{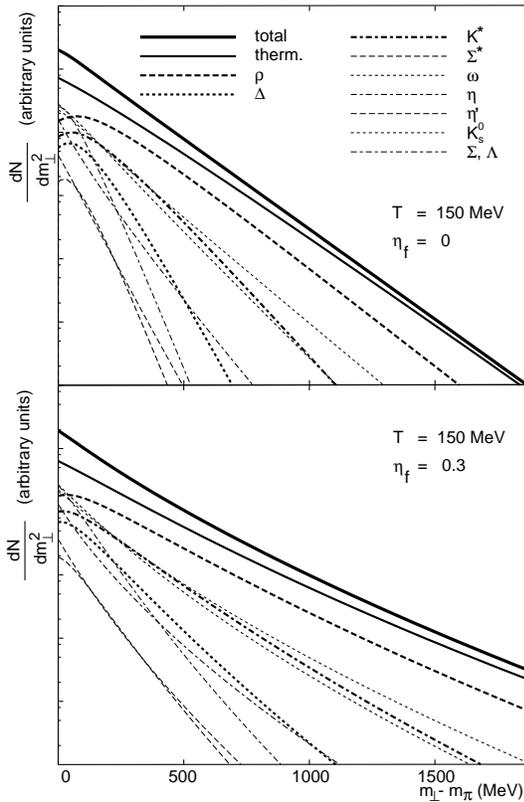}}
\caption{The single pion transverse mass spectrum for $T=150$ MeV 
and $\mu_B=\mu_S=0$. The overall normalization is arbitrary, the
relative normalizations of the various resonance contributions are fixed
by the assumption of thermal and chemical equilibrium. Upper panel: no
transverse flow, $\eta_f=0$; lower panel: $\eta_f=0.3$.
}\label{F2}
\end{figure}

For later reference we plot in Fig.~\ref{F2} the pion transverse
mass spectrum for the two sets of source parameters for which we 
compute two-particle correlations below. All resonance decay 
contributions are shown separately. The only 3-body decays are those
of the $\omega$, $\eta$ and $\eta'$ whose decay pions are seen to be 
particularly concentrated at small $p_\perp$. (A similar low-$p_\perp$
concentration occurs for pions from $K_S^0$ decays, due to the small
decay phase space in this particular 2-body decay.) Comparing the top panel 
(no transverse flow, $\eta_f=0$) with the bottom panel ($\eta_f=0.3$) 
one observes the well-known flattening of the transverse mass spectrum
by transverse radial flow \cite{SSH93,SH92,N82,LSH90}. The direct
pions reflect essentially an effective ``blueshifted'' temperature 
$T_{\rm eff} = T \sqrt{ {1 + \langle \beta_t \rangle \over 
1-\langle\beta_t\rangle}}$ \cite{LSH90}. But the heavier resonances,
in the region $P_\perp < M_r$, are affected much more strongly by
transverse flow since at small $P_\perp$ the flattening of the spectra
by flow is proportional to the particle rest mass \cite{N82,LSH90}. 
%
%
Fig.~\ref{F2}b shows that this effect on the parent resonances 
is also reflected in the spectra of the daughter pions, explaining 
the slight rise with $\eta_f$ of the resonance fractions at large 
$m_\perp$. This flattening of the transverse mass spectra by transverse 
flow, suggested in Refs.~\cite{LSH90,SSH93,SH92} as an explanation 
for the observed features of the single particle spectra from $^{28}$Si- 
and $^{32}$S-induced collisions at the AGS and SPS, seems to have been 
confirmed by recent collision experiments with very heavy 
ions (Au+Au at the AGS, Pb+Pb at the SPS, see contributions by 
Y. Akiba, R. Lacasse, Nu Xu and P. Jones at the recent Quark 
Matter '96 conference \cite{QM96}). One of the main goals of 
two-particle interferometry is to obtain an independent and 
more direct measure of the transverse expansion velocity at 
freeze-out, to confirm this picture and further discriminate against 
possible alternative explanations.
%
\begin{figure}[h]\epsfxsize=8.5cm 
\centerline{\epsfbox{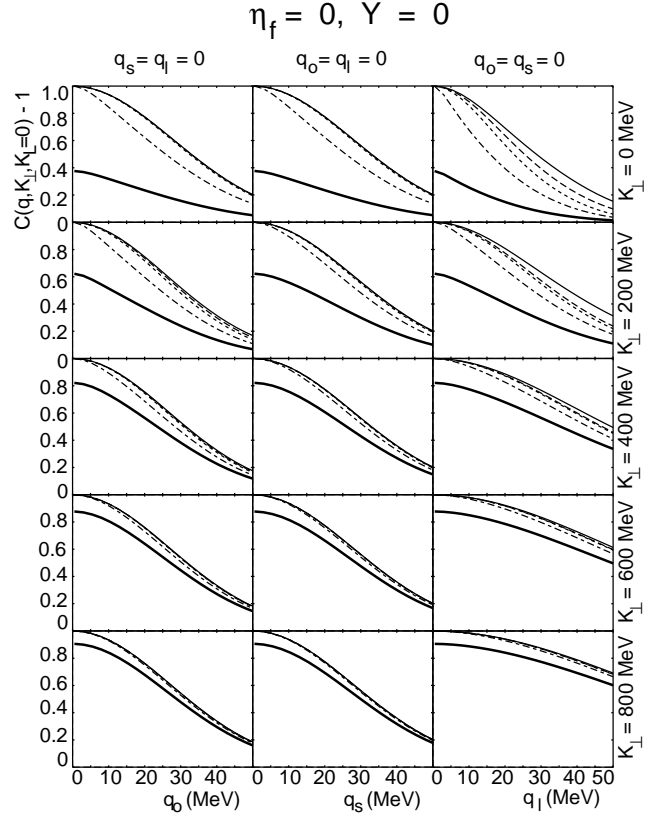}}
\caption{The two-particle correlator $C({\bf q},{\bf K})$ for $\pi^-$
pairs with pair rapidity $Y=0$ in the CMS. Each row of diagrams
corresponds to a different value for the transverse pair momentum 
$K_\perp$ ($K_\perp = 0, 200, 400, 600$ and 800 MeV from top to bottom).
Left column: the correlator in the outward direction at $q_s=q_l=0$.
Middle column: the correlator in the sideward direction at $q_o=q_l=0$.
Right column: the correlator in the longitudinal direction at $q_s=q_o=0$.
Source parameters as in Sec.~{\protect\ref{sec4}}, the transverse flow 
$\eta_f$ has been set to zero. \\
Here and in the following plots the different lines have the following 
meaning: {\em Thin solid line:} thermal pions only. {\em Long-dashed:}
including additionally $\rho$-decays. {\em Short-dashed:} including 
additionally all other shortlived resonances ($\Delta,K^*,\Sigma^*$, see 
Table~{\protect\ref{T1}}). {\em Dash-dotted:} adding also $\omega$-decays. 
{\em Thick solid line:} adding also all longlived resonances ($\eta,
\eta',K_S^0,\Sigma,\Lambda$, see Table~{\protect\ref{T1}}).
}\label{F3}
\end{figure}
%
\subsection{Two-particle correlations}
\label{sec5c}
 
\begin{figure}[h]\epsfxsize=8.5cm 
\centerline{\epsfbox{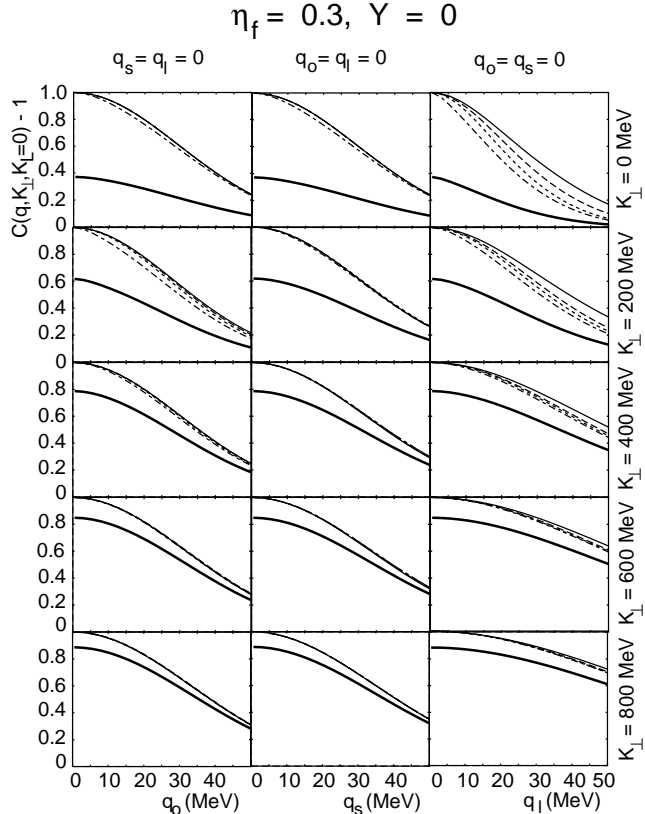}}
\caption{Same as Fig.~{\protect\ref{F3}}, but for nonzero transverse flow 
$\eta_f = 0.3$.
}\label{F4}
\end{figure}

In Figs.~\ref{F3} and \ref{F4} we plot the two-pion correlator 
$C({\bf q},{\bf K})$ in the three Cartesian directions of ${\bf q}$
for zero and non-zero transverse flow $\eta_f$, respectively. We use
the letter $Y$ to denote the rapidity {\em of the pair}, and $K_\perp$
($M_\perp$) for its transverse momentum (transverse mass). The pion
pairs in Figs.~\ref{F3} and \ref{F4} have pair rapidity $Y=0$ in the CMS, 
and transverse momenta ranging from 0 to 800 MeV/$c$ (top to bottom).
The correlation functions were calculated by numerically evaluating
Eq.~(\ref{3.3}) for the source parameters given in Sec.~\ref{sec4}.
%
%
Within each plot, the different lines show the effect of adding in 
Eq.~(\ref{3.3}) in the sum over decay channels $r$ successively more 
resonances (see Table~\ref{T1}): first the abundant $\rho$, then the 
other short-lived resonances, then the $\omega$ with its intermediate 
lifetime, and finally all the long-lived resonances. Comparing these 
plots row by row gives one a feeling for the $K_\perp$-dependence of 
the correlation function and the various resonance contributions. In 
the following two subsections we give a rough and general discussion 
of the main features of the correlator without and with transverse flow 
of the source, respectively, before proceeding to a quantitative analysis
in Sec.~\ref{sec6}.
 
\subsubsection{No transverse flow (Fig.~\protect\ref{F3})}
\label{sec5c1}
 
The direct thermal contribution leads to a correlation function 
with a nearly Gaussian shape in all directions $q_i$, $i = o,s,l$, and for
all pair momenta $K_\perp$. As $K_\perp$ increases, the correlator 
becomes rapidly wider in the longitudinal direction while in the two 
transverse directions the changes are hard to see and require a finer 
analysis (Sec.~\ref{sec6}). As more and more of the short-lived 
resonances are added, the width of the correlator becomes smaller, 
again with a larger effect in the longitudinal than in the two 
transverse directions. A much stronger effect is caused by the 
$\omega$-meson; now the narrowing of the correlation function is 
also clearly seen in the transverse directions, and the correlator 
becomes markedly non-Gaussian. As the long-lived 
resonances are added, the intercept $\lambda$ of the correlator at 
${\bf q}=0$ decreases below 1. This is a matter of $q$-resolution 
(we stop at $\vert{\bf q}\vert = 1$ MeV) -- the contribution from the
long-lived resonances is entirely concentrated in a $\delta$-function 
like structure near the origin, and with infinite resolution the 
correlator could be seen to actually reach the value 1 at ${\bf q}=0$.
This is, of course, an extreme deviation from Gaussian behaviour.
 
Long-lived resonances thus lead to apparently incomplete correlations,
$\lambda < 1$ \cite{Marb,CLZ96}. This effect becomes even stronger,
if the correlator is projected onto one particular $q$-direction by 
averaging over a finite window in the other directions where the 
correlator has already dropped below $\lambda$ \cite{MSH92}.
 
As the pion pair momentum $K_\perp$ increases, all resonance effects 
on the width and strength of the correlator are seen to decrease. This is
a direct consequence of the decreasing resonance fractions, see 
Fig.~\ref{F1}.
 
The above lifetime hierarchy of resonance effects can be understood 
in terms of the following simple picture:
 \begin{itemize}
 \item
{\it Short-lived resonances, $\Gamma > 30$} MeV. 
In the rest frame of the particle emitting fluid element these 
resonances decay very close to their production point, especially
if they are heavy and have only small thermal velocities.
This means that the emission function $S_{r\to\pi}$ of the daughter
pions has a very similar spatial structure as that of the parent 
resonance, $S_r^{\rm dir}$, although at a shifted momentum and shifted
in time by the lifetime of the resonance. As only $R_o$ and $R_l$ are 
sensitive to the lifetime of the source, the shift in time affects
the correlation function only in the outward and longitudinal directions.
The stronger effect on $R_l$ (which is obvious from the right column
in Fig.~\ref{F3}) is a consequence of the boost-invariant longitudinal 
expansion of our source: as the decay pions are emitted at a later
proper time $\tau$, and since the longitudinal length of homogeneity
increases with $\tau$ because the longitudinal velocity gradients 
decrease \cite{CSH95b}, the decay pions show a larger longitudinal 
homogeneity length than the direct pions. --
Since the Fourier transform of the direct emission function is rather 
Gaussian and the decay pions from short-lived resonances appear close 
to the emission point of the parent, they maintain the Gaussian features of 
the correlator.
 \item
{\it Long-lived resonances, $\Gamma \ll 1$} MeV. 
These are the $\eta$ and $\eta'$, with lifetimes $c\tau_\Gamma\approx
17.000$ and 1000 fm, respectively, and the weak decays of $K_S^0$ and 
the hyperons which on average propagate several cm. (The decays of 
$K_L^0$ and charged kaons are not included in our calculation 
because they are reconstructed in most experiments.) Even with 
thermal velocities these particles travel far outside the direct 
emission region before decaying, generating a daughter pion emission 
function $S_{r\to\pi}$ with a very large spatial support. The Fourier 
transform $\tilde{S}_{r\to\pi}(q,K)$ thus decays very rapidly for
$q\ne 0$, giving no contribution in the experimentally accessible 
region $q >1$ MeV. (This lower limit in $q$ arises from the finite 
two-track resolution in the experiments.) The decay pions do, however,
contribute to the single particle spectrum $\tilde{S}_{r\to\pi}(q=0,K)$
in the denominator and thus ``dilute'' the correlation. In this way
long-lived resonances decrease the correlation strength $\lambda$
without, however, affecting the shape of the correlator where it can 
be measured.
 \item
{\it Moderately long-lived resonances,} $1$ MeV $< \Gamma < 30$ MeV. 
There is only one such resonance, the $\omega$ meson. It is not 
sufficiently long-lived to escape detection in the correlator, and 
thus it does not affect the intercept parameter $\lambda$. Its lifetime 
is, however, long enough to cause a long exponential tail in
$S_{\omega\to\pi}(x,K)$. This seriously distorts the shape of the 
correlator and destroys its Gaussian form.
 \end{itemize}
%
 
\subsubsection{Non-zero transverse flow (Fig.~\protect\ref{F4})}
\label{sec5c2}
 
The main difference between Figs.~\ref{F3} and \ref{F4} is that
the effects from the short-lived resonances and the $\omega$ on the 
shape of the correlator are weaker. The primary reason for this behaviour
is that for the class of models (\ref{4.6}) the transverse size 
$R_t$ of the effective emission region for heavy resonances {\em shrinks} 
for nonzero transverse flow. In Gaussian saddle point approximation, 
this transverse size $R_t$ can be calculated from $S_r^{\rm dir}(x,P)$ 
in (\ref{4.6}) as
 \begin{equation}
    R_t = {R \over \sqrt{ 1 + (M_\perp/T) \eta_f^2}}\, .
 \label{5.9}
 \end{equation}
This is not accurate enough for quantitative studies \cite{WSH96} but
gives the correct tendency and right order of magnitude. Going like 
$\eta_f^2$ this effect is small, but it tends to increase the width 
of the correlator, counteracting the basic tendency of resonance
contributions to make the correlator narrower. For $\eta_f = 0.3$ the
two effects are seen to more or less balance each other in the sideward
correlator, leaving practically no trace of the shortlived resonances 
including the $\omega$. A similar effect is seen in the outward and 
longitudinal directions, but there the dominant lifetime effect
discussed above prevails.
 
Please note that none of the correlators shown in Figs.~\ref{F3},\ref{F4}
exhibits a ``volcanic'' (exponential or power law rather than Gaussian) 
shape as seen for the longitudinal correlators of Refs.~\cite{Marb}. 
We have not been able to trace the origin of this discrepancy; it may be
due to the different source (hydrodynamics with freeze-out along a sharp
hypersurface) used in Refs.~\cite{Marb}, but why this should manifest 
itself in this way is not obvious. From general arguments we would
expect at small $q$ a Gaussian behaviour with a curvature related to 
the longitudinal size of the effective pion source from $\omega$ 
decays; the longitudinal correlators in Refs.~\cite{Marb} seem to 
decay much more steeply for small $q$. We have checked our results 
with two independent programs, based on the formulae given in the 
Appendices.
 
\section{Extracting HBT-radii from the correlator}
\label{sec6}
 
Looking at Figs.~\ref{F3} and \ref{F4} it is clear that more quantitative
methods are needed to characterize the shape of the correlator.
For an interpretation of the correlator in terms of the space-time
structure of the source relatively small changes of its shape and its 
pair momentum dependence play an important role. One would therefore 
like to describe the key features of $C({\bf q}, {\bf K})$ by a small 
number of fit parameters which are sensitive to this space-time 
structure.  The usual procedure is to perform a Gaussian fit with the 
functions (\ref{2.6}) or (\ref{2.8}). As we will see this method runs 
into systematic problems if the 2-particle correlator does not have a 
perfect Gaussian shape, e.g. due to long-lived resonances. Not only do 
the functions (\ref{2.6}) or (\ref{2.8}) fail to give a good fit, but 
by not correctly accounting for the non-Gaussian features one throws 
away important space-time information contained in the resonance decay 
contributions to the correlator.  
 
In this Section we discuss several different Gaussian fitting procedures
which clearly demonstrate these difficulties. The main reason for
presenting this basically flawed approach is (i) that it is the method
mainly used so far in the experimental analysis and (ii) that the discussion
throws some light on how one should compare HBT radius parameters 
extracted by different groups using different procedures. After having 
understood the problems and the systematic uncertainties they generate 
we will then suggest a more reliable approach in the next Section which 
also accounts for non-Gaussian features in a quantitative way.
 
\subsection{Two-dimensional Gaussian fits to the correlator}
\label{sec6a}
 
We start by discussing 2-dimensional fits to $C({\bf q}, {\bf K})$ with 
two parameters $\lambda_i(K)$, $R_i(K)$ ($i=o,s,l$). We approximate the 
numerical function in the directions $q_i$ as follows:
  \begin{equation}
        \label{6.1}
        C(q_i,q_{k\ne i} = 0; {\bf K}) \approx
        1 + \lambda_i({\bf K})\, e^{-R_i^2({\bf K}){q_i}^2}\, , 
  \quad i=o,s,l\,.
  \end{equation}
The optimal parameters $\lambda_i({\bf K})$ and $R_i({\bf K})$ 
are determined by minimizing the following expression:
  \begin{equation}
        \sum_{\nu=1}^n \left( \ln C(q_i^\nu,q_{k\ne i} = 0; {\bf K}) 
        - \ln\lambda_i  + R_i^2 (q_i^\nu)^2 \right)^2 
        = {\rm min}\, .
        \label{6.2}
  \end{equation}
%
\begin{figure}[h]\epsfxsize=8.5cm 
\centerline{\epsfbox{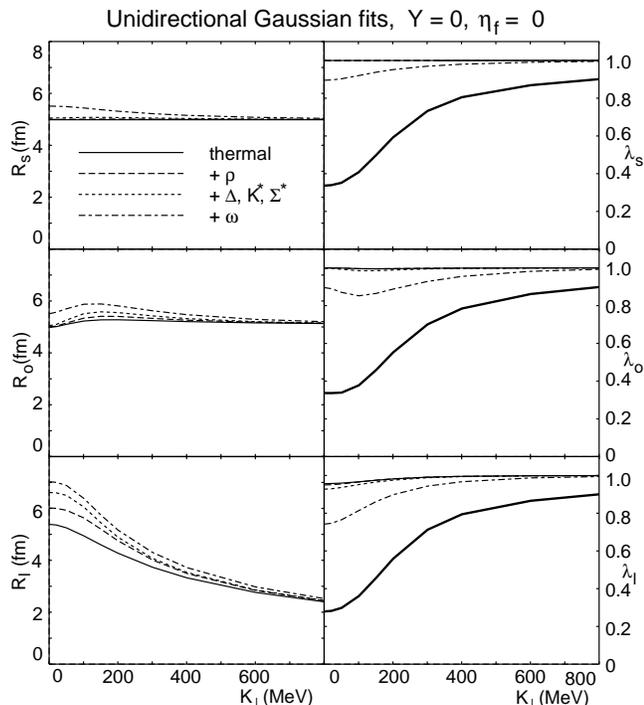}}
\caption{The Cartesian HBT radii $R_i$, $i = o,s,l$ and their
corresponding intercept parameters $\lambda_i$, extracted from
the correlator $C({\bf q},{\bf K})$ via 2-dimensional fits according
to Eq.~(\protect\ref{6.2}).
Shown are results at $Y=0$ as function of $K_\perp$ for $\eta_f=0$.
Top row: sideward direction. Middle row: outward direction. Bottom row:
longitudinal direction. The different lines indicate the effects of 
including various sets of resonances as described in Fig.~{\protect\ref{F3}}.
}\label{F5}
\end{figure}
%
The label $\nu$ runs over a set of $n$ equidistant values $q_i^\nu$ 
between $0$ and $50$ MeV for which  the correlator $C(q_i^\nu,{\bf K})$ 
was calculated numerically. Although the procedure (\ref{6.2}) is 
conceptually different from an experimental fitting procedure in that 
the function to be fitted is known exactly and the resulting optimal 
fit parameters thus don't have statistical error bars, they can still
vary systematically depending on the selection of the fit points 
$q_i^\nu$ and the minimization function (\ref{6.2}). These systematic
variations reflect the possible non-Gaussian features of the 
correlator, but not in a way that allows to easily quantify them.
As long as the deviations from Gaussian bahavior are small,
the extracted Gaussian fit parameters $R_i({\bf K})$ and 
$\lambda_i({\bf K})$ are expected to be useful for a simple 
characterization of the main features of the correlator.
 
\subsubsection{No transverse flow}
\label{sec6a1}
 
For the case $\eta_f=0$ the results from independent 2-dimensional
fits to the correlator in the ``side'' (top), ``out'' (middle), and
``long'' (bottom) directions are shown in Fig.~\ref{F5}. The left column 
shows the Cartesian HBT radii, the right column the associated
intercept parameters resulting from the fit, both as functions of $K_\perp$
at $Y=0$.
 
The fitted intercept parameters follow roughly the behaviour expected 
from Eq.~(\ref{3.12}) and Fig.~\ref{F1}. Upon closer inspection one
sees, however, that also some of the shorter-lived resonances,
in particular the inclusion of the $\omega$, have a significant lowering
effect on $\lambda$. These effects are different in the three Cartesian
directions and strongest in the longitudinal direction, where even 
without any resonance effects $\lambda<1$ at small $K_\perp$.
 
The deviations of the intercept parameter from unity reflect 
non-Gaussian features of the correlator. For short-lived resonances 
these are weak, except in the longitudinal $q_l$-direction where the
correlator has been known to show at small $K_\perp$ a somewhat 
steeper than Gaussian fall-off due to the rapid boost-invariant 
longitudinal expansion of the source \cite{WSH96}, even in the 
absence of resonance decays. The main non-Gaussian effects come from 
the $\omega$ and, of course, from the long-lived resonance. The latter 
affect, however, only $\lambda$ and not the HBT radii extracted from 
the Gaussian fit, while the $\omega$ also changes the radius parameters. 
 
The fit accomodates these non-Gaussian features by lowering the
intercept $\lambda$. As discussed in Sec.~\ref{sec5c} the main
origin of non-Gaussian effects due to resonances is the tail in the 
time-distribution of the decay pions. According to Eqs.~({2.7}) 
this is expected to affect $R_o$ and $R_l$, but not $R_s$. Eq.~({5.3}) 
tells us, however, that the ``out'' and ``side''-behaviour of the 
parent resonance distribution gets mixed in the pair distribution 
of the daughter pions, so some fraction of this effect propagates 
into the side-correlator of the decay pions. On the other hand there
remains the fact that, compared to $R_s$, in $R_o$ an additional 
lifetime effect comes in through the term $\beta_\perp^2 \langle 
\tilde t^2\rangle$ in (\ref{2.7b}); this contribution increases 
quadratically for small values of $K_\perp$, saturating above 
$K_\perp=m_\pi$ where $\beta_\perp \approx 1$. This explains very nicely 
the initial drop and subsequent rise of $\lambda$ in the outward 
direction, which is particularly prominent for the $\omega$ contribution.
 
Let us now turn our attention to the HBT radii in the left column of
Fig.~\ref{F5} and begin with a discussion of $R_s$. Its size remains 
essentially unaffected by the short-lived resonances with lifetimes 
of order $1$ fm/$c$, but the $\omega$ affects $R_s$. This effect dies 
out rapidly for increasing $M_\perp$ due to the decreasing 
$\omega$-fraction $f_\omega(K_\perp,0)$, but the resulting 
$M_\perp$-dependence of $R_s$ complicates the extraction of 
the transverse flow from it \cite{WSH96,WHTW96}.
 
The origin of the effect has already been qualitatively explained
in Sec.~\ref{sec5c} and above by referring to Eq.~(\ref{5.3}). A 
somewhat more quantitative estimate can be obtained by studying
the space-time variances of Sec.~\ref{sec3}, even though the 
discussion presented there makes it clear that this will provide 
only an upper estimate for the $\omega$-contribution to $R_s$.
Considering only the direct pions and those from $\omega$-decays
and calculating $R_s^2 = \langle y^2 \rangle$ according to (\ref{3.8})
we find
  \begin{equation}
        \label{6.3}
        {\langle{y^2}\rangle} = f_{\rm dir} {\langle{y^2}\rangle}_{\rm dir}
        + f_{\omega} {\langle{y^2}\rangle}_{\omega}
  \end{equation}
with
  \begin{mathletters}
  \label{6.4}
  \begin{eqnarray}
        {\langle{y^2}\rangle}_\omega &=&
        {{\int d^4x\, \sum_\pm \int_{\bf R} \int_0^\infty d\tau\Gamma 
        \, e^{-\tau\Gamma} y^2\, 
        S_\omega^{\rm dir}{\left(x-{P^\pm\over M}\tau,P^\pm\right)}}\over
        \int d^4x\, \sum_\pm \int_{\bf R} \int_0^\infty d\tau\Gamma 
        \, e^{-\tau\Gamma} 
        S_\omega^{\rm dir}{\left(x-{P^\pm\over M}\tau,P^\pm\right)}}
      \nonumber \\
        &=& R^2 + {\left({1\over \Gamma^2}\right)} f_{\rm corr}\, ,
        \label{6.4a} \\
        f_{\rm corr} &=& 2\, {{\int_{\bf R} {P_\perp^2\over M^2} \int d^4x
        S_\omega^{\rm dir}(x,P^\pm)\, \sin^2\Phi_\pm}\over
        {\int_{\bf R} \int d^4x\, S_\omega^{\rm dir}(x,P^\pm)}}\, .
        \label{6.4b}
  \end{eqnarray}
  \end{mathletters}
Using $f_{dir} + f_\omega = 1$ this yields
 \begin{equation}
 \label{6.5}
     {\langle{y^2}\rangle} = R^2 
     + f_{\rm corr}\, f_\omega\, {1\over \Gamma^2}\, .
 \end{equation}
This result can be explained as the effect of the $\omega$ propagating 
in the $y$-direction before decaying or, more formally, as the effect 
of the ``out''-``side''-mixing in the decay kinematics expressed by 
Eq.~(\ref{5.3}). Numerically, we determined the factor $f_{\rm corr} 
\approx 0.52$ at $K_\perp=0$ which leads to $f_{\rm corr}\cdot 
f_\omega \approx 0.1$ at the same point. Putting this together with 
the width of the $\omega$-resonance, ${1\over \Gamma} = 23.4$ fm, one 
obtains for the side variance $\sqrt{{\langle{y^2}\rangle} } = 8.9$ 
fm.  
 
This is obviously much larger than the $5.5$ fm extracted at $K_\perp 
= 0$, since the curvature (\ref{2.11}) does not coincide with the 
fitted width. For longer living resonances this discrepancy will, of 
course, be even larger. Another number to compare with is the half 
width $R_s^{\rm half}$ of the correlator $C(q_s)$ at $q_o{=}q_l{=}{\bf 
K}{=}0$, including all short-lived resonances plus the $\omega$. We 
find the hierarchy 
  \begin{equation}
        \label{6.6}
        \sqrt{ {\langle{y^2}\rangle} } \approx 8.9\text{ fm} >
        R_s^{\rm half} \approx 6.4 \text{ fm} >
        R_s \approx 5.5 \text{ fm}\, .
  \end{equation}
We conclude that estimates of resonance effects, based on space-time
variances like $\sqrt{ {\langle{y^2}\rangle} }$, as e.g. done in 
\cite{H96}, are quantitatively unreliable. The half width $R_s^{\rm 
half}$ is close to the result one would obtain from a Gaussian fit to 
the correlator {\em when the intercept $\lambda$ is fixed} to the
value of (\ref{3.12}) (as e.g. done, albeit simultaneously in all 
three $q$-directions, in Refs.~\cite{Marb,SOP96}). The difference to 
our procedure which lets $\lambda$ float is significant, and since 
$C$ at ${\bf q}=0$ is not experimentally accessible, a comparison of 
$R_s^{\rm half}$ with data \cite{SOP96} is clearly dangerous. The 
authors of Refs.~\cite{Marb} also find that at low $K_\perp$ resonance 
decays can increase the longitudinal HBT radius $R_l$ by up to 
a factor 2; in a Gaussian fit with floating $\lambda$ we never see
resonance induced increases in $R_l$ by more than 1.5 fm.
%
\begin{figure}[h]\epsfxsize=7cm 
\centerline{\epsfbox{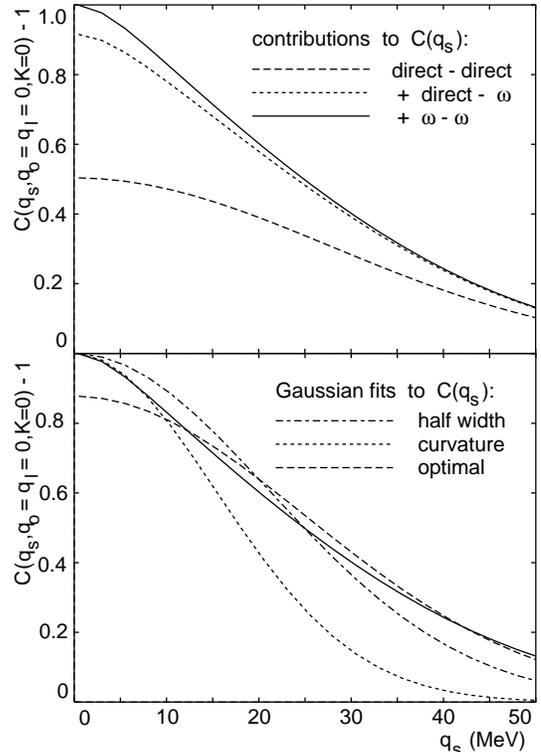}}
\caption{The correlator $C(q_s)$ at $q_o{=}q_l{=}{\bf K}{=}0$,
taking into account only direct pions and pions from $\omega$ decays.
The upper panel shows the three contributions according to 
Eq.~(\protect\ref{3.3}): direct-direct pairs (dashed), 
direct-direct + direct-$\omega$ (dotted), and all contributions (including
the $\omega-\omega$ term where both pions come from $\omega$ decays) 
(solid line).
The lower panel compares the same solid line to different Gaussians 
whose radius parameters correspond to the curvature at $q_s = 0$ 
(dotted line), the half width of $C(q_s)$ (dash-dotted line), 
and the optimal Gaussian fit according to Eq.~(\protect\ref{6.2}) 
(dashed line).
}\label{F6}
\end{figure}
%
We have compared the Gaussians corresponding to the numbers given in 
Eq.~(\ref{6.6}) with the true ``side''-correlator in the lower panel of 
Fig.~\ref{F6}. In the 
upper panel we show the three contributions to the correlator (see 
(\ref{3.3}) and (\ref{3.10})) coming from pairs of two directly 
emitted pions (dashed line), from pairs of one direct and one $\omega$ 
decay pion (difference between dotted and dashed lines), and from 
pairs where both pions come from $\omega$ decays (difference between 
solid and dotted lines). While it is obvious that the $\omega$ 
contributions are concentrated at lower $q_s$-values than the direct 
one, the tail from the mixed direct-$\omega$ contribution is still 
appreciable outside the half-point of the direct term near $q_s = 30$ 
MeV. It therefore appears impossible to cleanly separate the 
correlation function into ``core'' and ``halo'' contributions with 
different $q$-support \cite{CLZ96}. In particular, the recent 
suggestion by Cs\"org\H o \cite{Cs96} to extract the ``core'' radius 
by performing a Gaussian fit to the $q_s$-tail of the correlator, 
excluding the range $q_s<q_{\rm cut}$ where $q_{\rm cut} \sim 30$ MeV, 
is likely to run into systematic problems.  
 
We now turn to $R_o$. At $K_\perp=0$ the two transverse radius parameters 
$R_s$ and $R_o$ are equal by symmetry \cite{CNH95}, and all above 
considerations carry over to the ``out''-direction. At non-zero $K_\perp$, 
$R_o$ receives an additional contribution from the source lifetime as 
indicated by Eq.~(\ref{2.7b}). Although for the $\omega$ the use of this
expression is no longer quantitatively reliable, it gives the correct 
tendency. Short-lived resonances do not destroy the Gaussian shape of the
out-correlator, and for them Eq.~(\ref{2.7b}) (with Eq.~(\ref{3.8}))
can be used without restrictions. It is obvious that even the 
short-lived resonances contribute through their lifetime to the 
term $\beta_\perp^2 \langle \tilde t^2 \rangle$ in (\ref{2.7b}), 
strengthening the rise of $R_o$ in Fig.~\ref{F5} at small $K_\perp$. 
 
The strongest resonance effect is seen for $R_l$, which is affected 
even by the short-lived resonances. These effects disappear for
large $K_\perp$ due to the decreasing resonance fractions $f_r$, but
at small $K_\perp$ they are significant. Due to the existence of (weak)
non-Gaussian features already in the absence of resonances the
space-time variances are of limited use for a quantitative discussion of
the effects, and we leave the reader with the numerical results
shown in Fig.~\ref{F5}. A qualitative interpretation was given in 
Sec.~\ref{sec5c}.
 
\subsubsection{Non-zero transverse flow}
\label{sec6a2}
 
\begin{figure}[h]\epsfxsize=8.5cm 
\centerline{\epsfbox{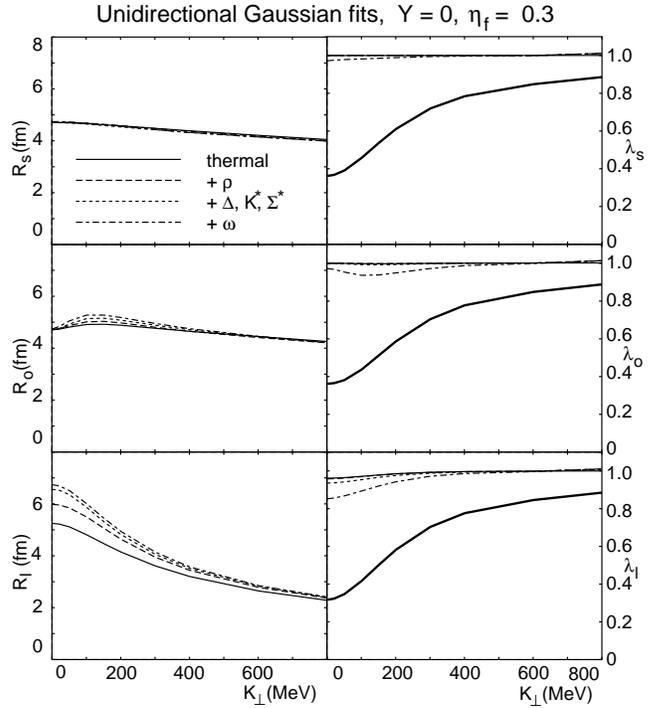}}
\caption{Same as Fig.~\protect\ref{F5}, but for non-zero transverse flow
$\eta_f = 0.3$.
}\label{F7}
\end{figure}

In Fig.~\ref{F7} we show the parameters $\lambda_i$, $R_i$ obtained 
from the 2-dimensional fit (\ref{6.1}), (\ref{6.2}) for the case of 
non-zero transverse expansion with $\eta_f=0.3$.  Comparing with 
Fig.~\ref{F5} one sees that the effect of the resonances on the radius 
on the HBT radius parameters are weaker, and that correspondingly the 
non-Gaussian effects caused by the short-lived resonances and the 
$\omega$ (which lead to deviations of the intercept parameter 
$\lambda$ from unity) are less pronounced. In fact, the only remaining 
effects of these resonances on the HBT radii come from the terms $\sim 
\langle \tilde t^2 \rangle$ in $R_o$ and $R_l$ (see (\ref{2.7})) and 
are due to the additional contribution to the particle emission 
duration from the resonance lifetimes. The geometrical effect of 
resonance propagation away from the direct source, described by the 
second term in (\ref{6.5}), has disappeared, even for the $\omega$.  
The reason was already discussed in Sec.~\ref{sec5c}, Eq.~(\ref{5.9}): 
due to transverse flow the transverse size of the emission region for 
heavy resonances is smaller than that of the direct pions, and since 
they don't live very long they usually don't make it outside the 
source of direct pions before decaying. Thus they don't lead to an 
increase of the spatial source size.  
%
%
As a consequence, the decrease of $R_s$ with increasing $K_\perp$, 
which is characteristic for transverse collective flow of the source 
\cite{WSH96,WHTW96}, is no longer modified by the pions from $\omega$ 
decays. This is, of course, the ideal situation one might hope for in 
order to extract quantitative dynamical information from HBT data. 
Unfortunately, the problem remains that, if the measurement finds
a (not too strong) $M_\perp$-dependence of $R_s$, it could still be 
due to either weak transverse flow without resonance contaminations
as in Fig.~\ref{F7}, or to $\omega$-decay contributions in the absence
of transverse flow as in Fig.~\ref{F5}. (Other mechanisms like 
transverse temperature gradients might also create an $M_\perp$-dependence 
\cite{CL94,CSH95b}.) We must find a more quantitative analysis tool 
which allows us to tell whether the $M_\perp$-dependence of $R_s$ is 
associated with $\omega$-decays or not.
 
\subsection{Five-dimensional Cartesian Gaussian fits to the correlator}
\label{sec6b}
 
Before approaching this task in the next Section, we will now also 
discuss some generic features of multi-dimensional Gaussian fits to the
exact correlator where all HBT parameters and the correlation strength
are determined simultaneously. This is clearly desirable in order to
avoid the problem of having three different correlation strengths in
the three Cartesian directions, as happens when the three radii $R_s$,
$R_o$, and $R_l$ are determined by separate 2-dimensional fits according
to (\ref{6.1}), because such a result is clearly unphysical. It is also
necessary for the determination of the cross-term $R_{ol}$ and for
a fit with the YKP parametrization (\ref{2.8}).
%
\begin{figure}[h]\epsfxsize=8.5cm 
\centerline{\epsfbox{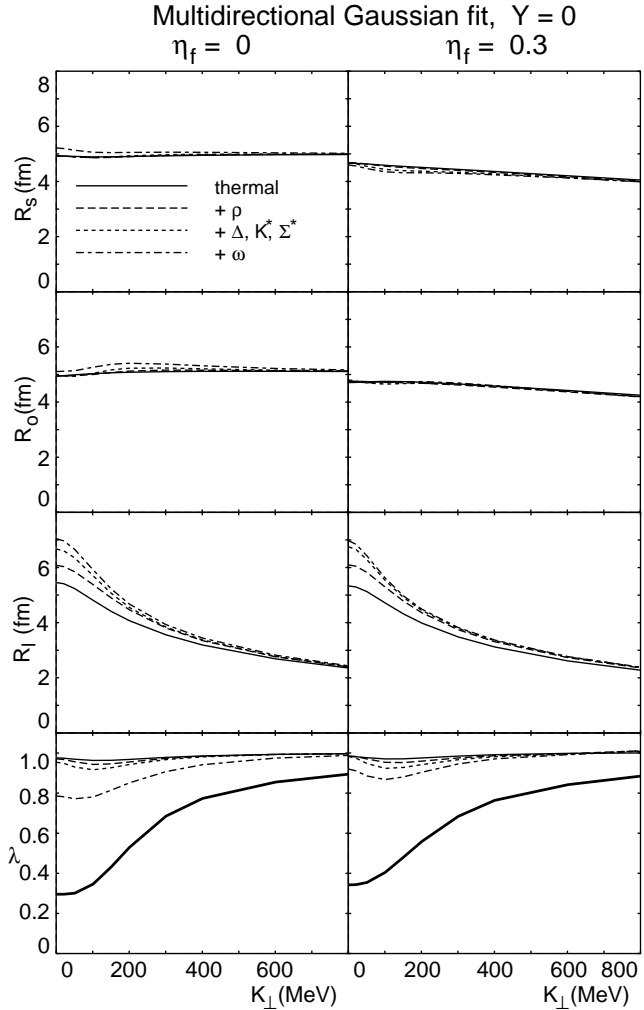}}
\caption{The Cartesian HBT radii $R_o$, $R_s$, $R_l$, and
the intercept $\lambda$, obtained from the 5-dimensional fit 
(\protect\ref{6.7}) to $C({\bf q},{\bf K})$, as functions of 
$K_\perp$ at $Y=0$. $R_{ol}$ is not shown since at $Y=0$ it vanishes
due to symmetry. Left column: no transverse flow, 
$\eta_f=0$. Right column: $\eta_f=0.3$.
}\label{F8}
\end{figure}
%
In this Subsection we extract the Cartesian parameters 
$R_o$, $R_s$, $R_l$, $R_{ol}$ and $\lambda$ from a 5-dimensional
fit which minimizes the expression
  \begin{eqnarray}
    &&\sum_{\nu=1}^n \left[ \ln C({\bf q}^\nu,{\bf K}) - \ln\lambda
          + R_o^2\, (q_o^\nu)^2
          + R_s^2\, (q_s^\nu)^2
          + R_l^2\, (q_l^\nu)^2 \right.
        \nonumber \\
        && \qquad  + 2\, R_{ol}^2\, q_o^\nu\, q_l^\nu \Big)^2 = \text{min}.
  \label{6.7}
  \end{eqnarray}
The label $\nu$ again runs over $n$ fit points ${\bf q}^\nu$ which 
were chosen to lie at equal distances between 0 and 50 MeV along the 
three Cartesian $q$-axes and along the two diagonals $(q_s=0,q_o=q_l)$ 
and $(q_s=0,q_o=-q_l)$. This is, of course, different from a typical 
experimental ${\bf q}$-distribution. They were selected to economize 
in the number of fit points where the exact correlator had to be 
computed.  
 
The results of the fit (\ref{6.7}) are shown in Fig.~\ref{F8}, again 
for $\eta_f=0$ and $\eta_f=0.3$ at midrapidity $Y=0$. Let us first 
look at the incercept parameter $\lambda$. Comparing with Figs.~\ref{F5}, 
\ref{F7} we see that the $\lambda$-value from the 5-dimensional fit 
lies somewhere between the three different values obtained in the 
2-dimensional fits. As before it reflects the deviations of the 
correlator from a Gaussian. Since such deviations exist in the 
$q_l$-direction even without resonances, due to strong longitudinal 
expansion, $\lambda$ slightly deviates from 1 even in the absence of 
resonance decays.  
 
The need for the fit to compromise on a unique intercept parameter
affects the optimum values for the HBT radius parameters. For a fixed
correlation function, a decrease of $\lambda$ leads automatically also
to a smaller Gaussian radius as found by the fit. Since the compromise
value for $\lambda$ lies above the value $\lambda_l$, but below the 
values $\lambda_s$ and $\lambda_o$ from the 2-d fits, $R_l$ increases 
and $R_s$, $R_o$ decrease in the 5-d fit relative to the 2-d fit 
values. (This effect is hardly visible if resonance decays are 
switched off but becomes stronger as the resonance contributions (with 
their non-Gaussian effects) are added.) The net result is that even in 
the absence of transverse flow now the resonance effects on $R_s$ and 
on its $M_\perp$-dependence appear quite weak. Even the resonance 
contribution to the lifetime effect in $R_o$ (the quadratic rise at 
small $K_\perp$) becomes less pronounced.  
 
\begin{figure}[h]\epsfxsize=8.5cm 
\centerline{\epsfbox{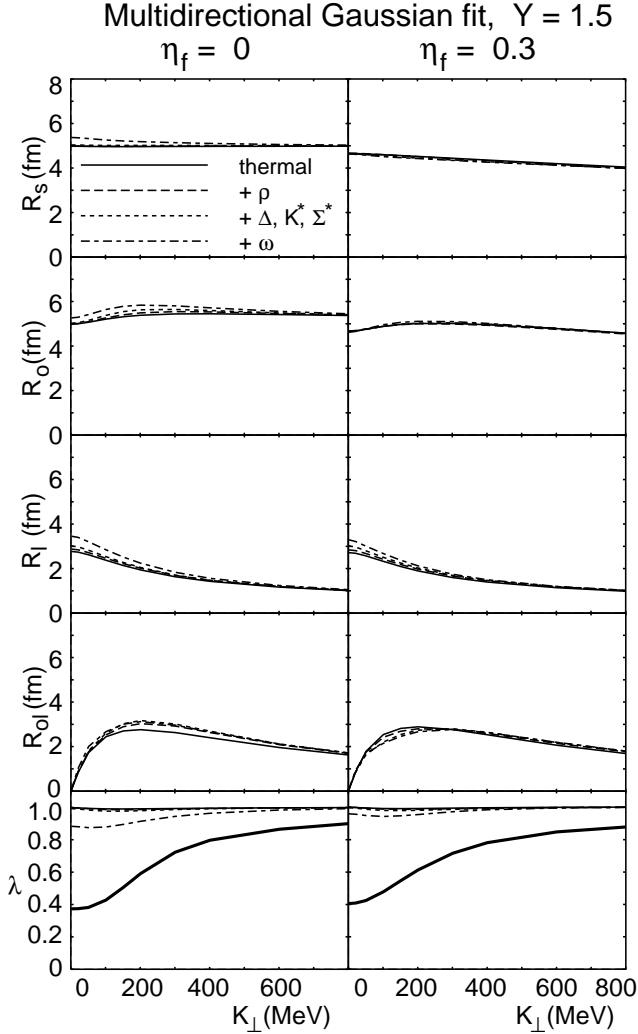}}
\caption{Same as Fig.~\protect\ref{F8}, but for forward rapidity $Y=1.5$.
Now also $R_{ol}$ is non-zero.
}\label{F9}
\end{figure}

For completeness we show in Fig.~\ref{F9} also results at forward
pair rapidity $Y=1.5$. In this case the fit gives, of course, a
non-vanishing cross-term $R_{ol}$ with the expected $K_\perp$-behaviour
\cite{TH96}. It is affected by resonances essentially at the same level 
as $R_o$. The only other qualitative difference is the much smaller value 
of $R_l$ relative to $Y=0$; this is an effect of Lorentz contraction.
 
It must be stressed that the differences between this Subsection
and the previous one are purely due to fit systematics. Depending on 
how the exact correlator is fitted to a Gaussian the extracted Gaussian 
radii show significant differences. Since in the experiments the intercept
parameter cannot be directly measured and must be fitted simultaneously
with the HBT radii, we adopted the same procedure and let $\lambda$ float
in the fit. Schlei \cite{Marb,SOP96}, on the other hand, in 
his Gaussian fits has always fixed $\lambda$ at the value given by 
Eq.~(\ref{3.12}). In the presence of non-Gaussian effects due to 
resonance decays our fits give smaller $\lambda$'s and, therefore,
smaller HBT radii than his fit. This explains why the resonance effects 
on the transverse radii and their $K_\perp$-dependence were found to 
be much stronger in Refs.~\cite{Marb,SOP96} than in our work 
here. According to the second inequality in (\ref{6.6}) the difference 
in $R_s$ is about 1 fm, in good agreement with his compared to our results.
%
%
\subsection{Five-dimensional Gaussian YKP fits to the correlator}
\label{sec6c}
 
The extraction of the Yano-Koonin velocity from a fit according to 
Eq.~(\ref{2.8}) is a non-linear problem. To maintain the simplicity
of a least-square fit with linear fit parameters we have reformulated
the YKP fit problem as follows: We rewrite (\ref{2.8}) in the form
 \begin{eqnarray}
   C({\bf K},{\bf q}) &=& 1 + \lambda_{_{\rm YKP}}({\bf K})
   \exp\left[ - R_\perp^2({\bf K})\, q_\perp^2 
              - R_{33}^2({\bf K}) q_l^2 \right.
            \nonumber \\
            && \qquad \left.
              - R_{00}^2({\bf K}) (q^0)^2
              + 2 R_{03}^2({\bf K}) q^0 q_l
                \right]\,  ,
 \label{6.8}
 \end{eqnarray}
with
 \begin{mathletters}
 \label{6.9}
 \begin{eqnarray}
 \label{6.9a}
   v &=& {1\over 2D} \left( 1 - \sqrt{1 - \left({4D^2}\right)^2}
                       \right) \, ,
 \\
 \label{6.9b}
   R_0^2 &=& { {R_{00}^2 - v^2\, R_{33}^2} \over {1 + v^2}} \, ,
 \\
 \label{6.9c}
   R_\parallel^2 &=& { {R_{33}^2 - v^2\, R_{00}^2} \over {1 + v^2}} \, ,
 \\
 \label{6.9d}
   D &=&{ R_{03}^2\over {R_{00}^2 + R_{33}^2}}\, .
 \end{eqnarray}
 \end{mathletters}
We then proceed as with the Cartesian parametrization in 
Sec.~\ref{sec6b}, using the same set of fit points as before, but 
expressing them through their components $q_\perp,q_l$, and $q^0$, in 
order to determine $\lambda_{_{\rm YKP}}, R_\perp, R_{33}, R_{00}$, 
and $R_{03}$.  Finally we solve Eqs.~(\ref{6.9}) for the YKP 
parameters.  
%
\begin{figure}[h]\epsfxsize=8.5cm 
\centerline{\epsfbox{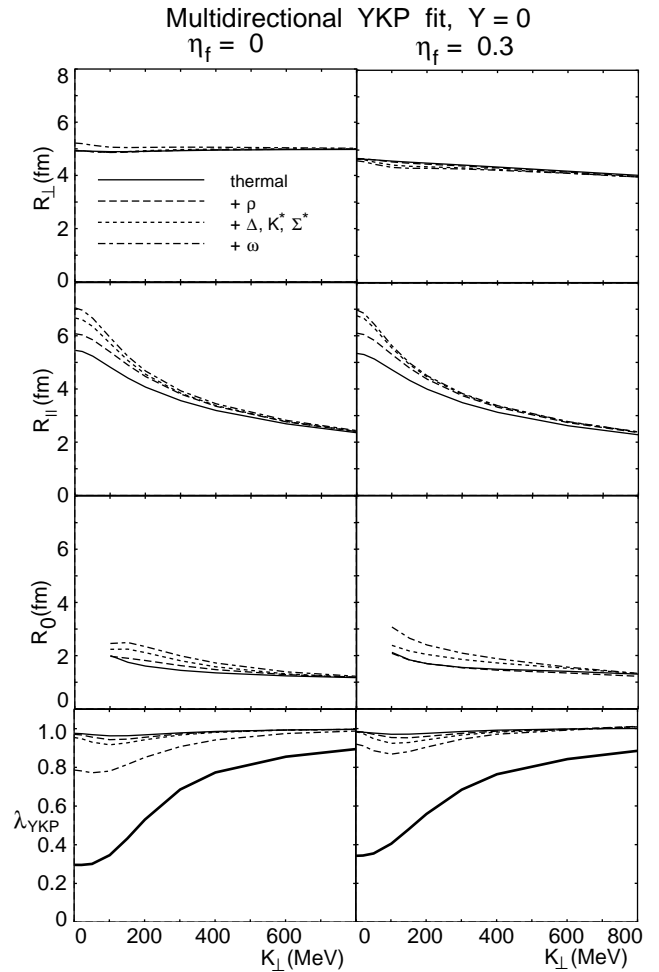}}
\caption{From top to bottom: the YKP fit parameters $R_\perp$, $R_0$, 
$R_\parallel$, and $\lambda_{_{\rm YKP}}$ as functions of $K_\perp$ at 
mid rapidity $Y = 0$. Left column: no transverse flow, $\eta_f = 0$.
Right column: $\eta_f = 0.3$. At $Y=0$ the Yano-Koonin velocity $v$
vanishes exactly.
}\label{F10}
\end{figure}
%
However, the one-to-one correspondence between the YKP and Cartesian 
radius parameters  does not imply that in a fit to experimental data 
both sets of fit parameters can be determined with similar accuracy.  
At mid rapidity, for instance, where $q^0 = \beta_\perp\, q_o$, the 
YKP fit becomes for small transverse pair momentum increasingly 
insensitive to $R_0$, since $q^0 \to 0$ for $\beta_\perp \to 0$.  As a 
result, in the space of YKP fit parameters the confidence region for 
one standard deviation is very elongated in $R_0$.  The actual fit 
value of $R_0$ thus develops a strong sensitivity to relatively small 
systematic deviations of the correlator from a Gaussian shape. Since 
the procedure (\ref{6.7}) adopted here does not allow to associate 
errors to the extracted fit values, we present in Figs.~\ref{F10} and 
\ref{F11} the results only for sufficiently large values of $K_\perp$ 
where such systematic effects were found to be small.  
 
For $Y=0$ ($\beta_l = 0$, Fig.~\ref{F10}) the systematic uncertainty at
small values of $K_\perp$ ($\beta_\perp$) affects only $R_0$, according
to the arguments presented above. For $K_\perp < 100$ MeV, we found 
that the $R_0$ value extracted from the Gaussian fit develops a strong
dependence on the choice of the fit points ${\bf q}^{\nu}$ while this
problem disappears at larger values of $K_\perp$. For $Y=1.5$,
($\beta_l \not= 0$, Fig.~\ref{F11}), similar systematic uncertainties
at small $K_\perp$ affect also $R_\parallel$ and $Y_{YK}$. Accordingly,
the corresponding curves in Fig.~\ref{F11} have been cut off at small
$K_\perp$.
 
The intercept parameters $\lambda_{_{\rm YKP}}$ extracted from the fit 
to (\ref{6.8}) essentially coincide with those from the 5-dimensional
Cartesian fit. This is expected since in both fits the same set of 
fit points was used. Also, the results for $R_\perp$ compare very well 
with $R_s$ in the Cartesian fit. For a Gaussian correlator the formalism 
of space-time variances says $R_\perp^2 = R_s^2 = {\langle{y^2}\rangle}$. 
The equality $R_\perp=R_s$ remains essentially unaffected by the
non-Gaussian features of the correlator in the presence of resonance 
decays.
  
The longitudinal YKP parameter $R_\parallel$ is affected by resonance 
decays roughly in the same way as $R_l$ in the Cartesian fit at $Y=0$.
This is expected because at $Y=0$ the two parameters are again identical 
on the level of space-time variances, see Eqs.~(\ref{2.7c}), (\ref{2.10b}).
There is no drastic change for $R_\parallel$ as one goes from $Y=0$ to
$Y=1.5$: All values (with and without resonances) decrease somewhat,
because one approaches the forward end of the source, and the longitudinal
homogeneity region thus shrinks a bit.
 
The most significant resonance contribution is seen in the lifetime 
parameter $R_0$. This agrees with our arguments that the dominant
effect from resonances on the correlation function arises from their
finite lifetime.
 
At $Y=0$ the Yano-Koonin velocity $v$ vanishes \cite{HTWW96,WHTW96}. 
This is reproduced by the fit. At forward rapidity $v$ is non-zero.
In the fourth row of Fig.~\ref{F11} we plot the Yano-Koonin rapidity 
$Y_{_{\rm YK}} = {1\over 2} \ln{\left({ {1 + v}\over {1 - v} }\right)}$
as a function of the transverse pair momentum. For longitudinally 
boost invariant sources, the YK rapidity is known to coincide with the
pair rapidity, $Y_{_{\rm YK}}(K_\perp,Y)=Y$. For the class of models of 
Sec.~\ref{sec4} with longitudinally boost-invariant flow previous 
studies without resonance decay contributions gave a linear relation 
between the two quantities, $Y_{_{\rm YK}}(K_\perp,Y)= c(K_\perp)\, Y$. 
The $Y$-dependence $Y_{_{\rm YK}}$ provides direct experimental access 
to the longitudinal expansion of the source. For thermalized models the 
proportionality constant $c(K_\perp)$ slowly approaches unity from 
below as $K_\perp$ increases \cite{WHTW96}. This is clearly seen in 
Fig.~\ref{F11} which also shows that resonance decay contributions
have a negligible influence on this relation.
%
\begin{figure}[h]\epsfxsize=8.5cm 
\centerline{\epsfbox{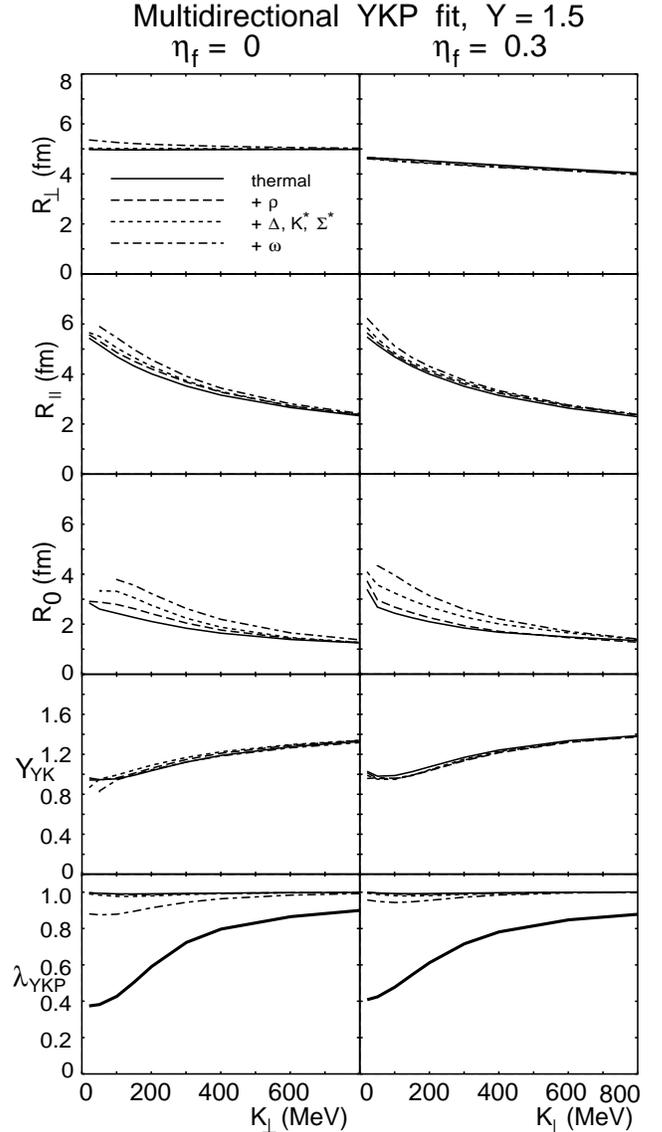}}
\caption{Same as Fig.~\protect\ref{F10}, but for forward rapidity $Y = 1.5$.
The additional fourth row now shows the Yano-Koonin rapidity at function
of $K_\perp$.
}\label{F11}
\end{figure}
%
\section{$\lowercase{q}$-variances of the correlator}
\label{sec7}
 
We have seen that resonances, in particular the $\omega$ with its 
intermediate lifetime, create appreciable non-Gaussian effects in the 
two-pion correlator, and that these deviations from a Gaussian shape 
can thus contain additional information about the space-time 
distribution of the source and its physical origin. They also have 
negative effects on the extraction of HBT radius parameters from 
Gaussian fits and affect their $K_\perp$-dependence in a way which,
within the framework of Gaussian fits, is difficult to quantify and 
to control systematically.
 
In this Section we therefore study an alternative approach. We suggest
to extract the HBT radius parameters and quantify the deviations from
Gaussian behaviour by studying the normalized second and fourth order
$q$-moments of the correlator $C({\bf q,K})$. We first develop the 
necessary formalism and then apply it to the correlation functions 
calculated from our class of source models.
 
\subsection{General formalism}
\label{sec7a}
 
According to Sec.~\ref{sec2}, the most general Gaussian ansatz for the 
correlator is 
   \begin{equation}
     C({\bf q},{\bf K}) = 1 + \lambda({\bf K})\, 
     \exp\left[ -\sum_{i,j=1}^3 q_i\, D_{ij}({\bf K})\, q_j \right]\, ,
     \label{7.1}
   \end{equation}
where the $q_i$ are the three independent relative momentum components
obtained after resolving the on-shell constraint $q^0 = \bbox{\beta}
\cdot {\bf q}$. For such a Gaussian correlator, the HBT parameters
$D_{ij}({\bf K})$ can be obtained by either fitting the various widths
of the correlator as done in Sec.~\ref{sec6}, or by computing the integrals
   \begin{equation}
     \langle\!\langle q_i\, q_j \rangle\!\rangle \equiv
     {\int d^3q\, q_i\, q_j\, [ C({\bf q},{\bf K}) - 1 ]
      \over 
      \int d^3q\, [ C({\bf q},{\bf K}) - 1] }
     = {1\over 2} \left(D^{-1}({\bf K})\right)_{ij}
   \label{7.2}
   \end{equation}
and inverting the resulting matrix of second order $q$-moments. 
 
For a non-Gaussian correlator we may {\em define} the HBT radius
parameters in terms of these ``$q$-variances'': having determined
the matrix $D({\bf K})$ by inverting the matrix 
$\langle\!\langle q \otimes q \rangle\!\rangle ({\bf K})$ of 
$q$-variances, we define
  \begin{equation}
  \label{7.3}
        \left(
        \begin{array}{ccc}
                R_s^2    & R_{os}^2 & R_{ls}^2 \\
                R_{os}^2 & R_o^2    & R_{ol}^2 \\
                R_{ls}^2 & R_{ol}^2 & R_l^2
        \end{array}\right)
     \equiv
        \left(
        \begin{array}{ccc}
                D_{ss}   & D_{os}   & D_{ls} \\
                D_{os}   & D_{oo}   & D_{ol} \\
                D_{ls}   & D_{ol}   & D_{ll}
        \end{array}\right)
  \end{equation}
when $q_s,q_o,q_l$ are used as independent coordinates, and
  \begin{equation}
  \label{7.4}
        \left(
        \begin{array}{ccc}
                R_\perp^2 & 0         & 0         \\
                0         & R_{33}^2  & -R_{03}^2 \\
                0         & -R_{03}^2 & R_{00}^2
        \end{array}\right)
     \equiv
        \left(
        \begin{array}{ccc}
                D_{\perp\perp}  & 0        & 0      \\
                0               & D_{33}   & D_{03} \\
                0               & D_{03}   & D_{00}
        \end{array}\right)
  \end{equation}
if one uses instead $q_\perp, q_l, q^0$ as independent variables. 
Eq.~(\ref{7.3}) corresponds to the Cartesian parametrization (\ref{2.6}),
generalized to systems without azimuthal symmetry by allowing for
non-vanishing ``side-out'' and ``side-long'' cross terms. Eq.~(\ref{7.4})
corresponds to the YKP parametrization (\ref{2.8}) which applies only to
azimuthally symmetric systems, and the zeroes in the matrices on the left 
and right hand side reflect this symmetry.
 
Similarly, the intercept parameter can be defined in terms of the 
$q$-variances and the zeroth order $q$-moment as
  \begin{equation}
        \label{7.5}
        \lambda({\bf K}) = \pi^{-3/2}\,\sqrt{\det D({\bf K})}
        \int d^3q\, [ C({\bf q},{\bf K}) - 1] \, ,
  \end{equation}
which reproduces the correct value for Gaussian correlators of type 
(\ref{7.1}).
 
The deviations from Gaussian behaviour in the correlator are then 
related to higher order $q$-moments. A general discussion, including 
their derivation from a generating functional from which the full 
correlator can be reconstructed, is given in Ref.\cite{WH97}.
Since $C({\bf q,K})$ is symmetric with respect to interchange of the 
particle momenta $p_1$ and $p_2$ and therefore even under ${\bf q} \to 
- {\bf q}$, all odd $q$-moments vanish. The first non-Gaussian 
contributions thus show up in the fourth order moments.
 
Application of the method of $q$-moments thus generally requires at 
least an inversion of the matrix (\ref{7.2}) for the determination of the HBT 
radius parameters and a discussion of the 4-dimensional tensor of
fourth order moments for the non-Gaussian aspects. A complete such 
analysis in three-dimensional $q$-space will be postponed to a 
future publication. Here we will perform a unidirectional 
analysis, where these technical complications do not arise, and compare 
it with the unidirectional Gaussian fits of Sec.~\ref{sec6a}.
 
We thus consider the correlator along one of the three axes $q_i$
($i=s,o,l$ or $i=\perp,l,0$) which we denote by $\tilde C(q_i)$, 
suppressing for simplicity the ${\bf K}$-dependence:
 \begin{equation}
 \label{7.6}
    \tilde C(q_i) \equiv C(q_i, q_{j\ne i}=0,{\bf K})\, .
 \end{equation}
The HBT radius parameter in direction $i$ and the corresponding intercept
parameter are then defined as
  \begin{mathletters}
  \label{7.7}
  \begin{eqnarray}
  \label{7.7a}
     R_i^2 &=& {1\over 2\,\langle\!\langle q_i^2 \rangle\!\rangle}\, ,
  \\
  \label{7.7b}
     \langle\!\langle q_i^2 \rangle\!\rangle
     &=& {\int dq_i\, q_i^2\, [\tilde C(q_i)-1] \over
          \int dq_i\, [\tilde C(q_i)-1] }\, ,
  \\
  \label{7.7c}
     \lambda_i &=& (R_i/\sqrt{\pi})
     \int dq_i\, [\tilde C(q_i)-1] \, .
  \end{eqnarray}
  \end{mathletters}

To extract the moments ${\langle\!\langle{q_i^n}\rangle\!\rangle}$ 
from data one replaces (\ref{7.7b}) by a ratio of sums over bins in the 
$q_i$-direction. The higher the order $n$ of the $q$-moment, the 
more sensitive are the extracted values to statistical and systematic
uncertainties in the region of large $q_i$. First investigations with 
event samples generated by the VENUS event generator indicate that the 
current precision of the data in the Pb-beam experiments at the CERN 
SPS permits to determine the second and fourth order $q$-moments. 
Accordingly, we restrict our discussion of non-Gaussian 
features to the ``kurtosis''
  \begin{equation}
  \label{7.8}
    \Delta_i = { \langle\!\langle q_i^4 \rangle\!\rangle
                 \over
                3\, \langle\!\langle q_i^2 \rangle\!\rangle^2} - 1\, .
  \end{equation}
In the following Sec.~\ref{sec7b} we will study the $K_\perp$-dependence
of the HBT radius parameters, the intercept and the kurtosis as defined by
Eqs.~(\ref{7.7}) and (\ref{7.8}).
 
\subsection{Unidirectional results for the $q$-moments}
\label{sec7b}
 
In this Subsection we present a numerical analysis of the correlation
functions computed in Sec.~\ref{sec5} in terms of their $q$-moments
along the three Cartesian directions, and give a comparison with the
unidirectional Gaussian fits presented in Sec.~\ref{sec6a}.
 
\begin{figure}[h]\epsfxsize=8.5cm 
\centerline{\epsfbox{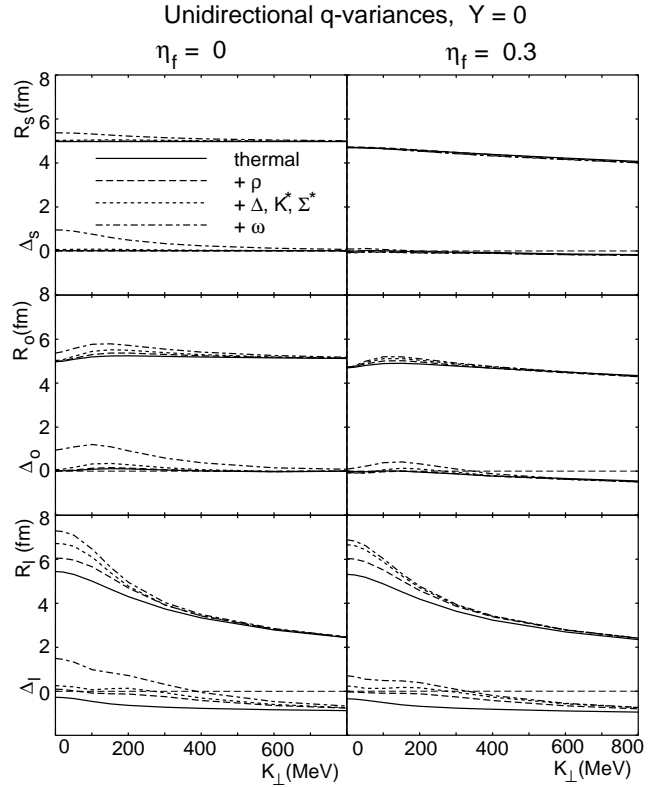}}
\caption{1-dimensional $q$-variances according to Sec.~\protect\ref{sec7a}.
The plots show for the three Cartesian directions $i = s,o,l$
the radius parameters $R_i(K_\perp)$ defined by Eq.~(\protect\ref{7.7a}) 
(upper set of curves in each panel) and the kurtosis $\Delta_i(K_\perp)$ 
defined by Eq.~(\protect\ref{7.7a}) (lower set of curves in each panel).
The radii are given in fm, the kurtosis in dimensionless units on the 
same scale. Left column: no transverse flow, $\eta_f=0$. Right column:
$\eta_f=0.3$. The pion pairs have rapidity $Y=0$ in the CMS.
}\label{F12}
\end{figure}

Fig.~\ref{F12} shows the HBT radii (\ref{7.7a}) and the kurtosis 
(\ref{7.8}) along the ``side''-, ``out'-, and ``long''-axes (from 
top to bottom). The left and right column of plots correspond to zero
and non-zero transverse flow of the source, respectively. In each panel 
we plot as the upper set of curves the HBT radius parameter $R_i$ in fm,
with different line symbols denoting the effects of including various 
sets of resonances as before. They should be compared with the lines
shown in the left columns of Figs.~\ref{F5} and \ref{F7}, respectively.
The lower set of lines (clustered around values near 0) denote the 
corresponding kurtosis $\Delta_i$ in dimensionless units. These contain
the lowest order information on the non-Gaussian features of the
numerically computed correlation function.
  
The comparison of the HBT radius parameters defined via the $q$-variances
of the correlator with those from the Gaussian fit (\ref{6.2}) shows
a remarkable agreement. As stressed above, in the presence of non-Gaussian
features in the correlator, the only well-defined definition of the HBT
radii is provided by the $q$-variances (\ref{7.7a}), while the Gaussian
fit results have possibly severe systematic uncertainties related to the
details of the fit procedure. The agreement between the corresponding 
curves in Figs.~\ref{F5}, \ref{F7}, and \ref{F12} indicates that we were
``lucky'' with our choice of fit prescription in Sec.~\ref{sec6a}. An
essential reason for the good agreement was our decision to let the 
intercept parameter $\lambda$ float in Eq.~(\ref{6.2}), i.e. to perform
a 2-dimensional rather than a 1-dimensional fit as in 
Refs.~\cite{Marb}. The discrepancy between the HBT radii shown in
those papers and those shown in Fig.~\ref{F12} thus simply reflect
the systematic uncertainties of extracting a Gaussian width parameter 
from a non-Gaussian correlator. In view of these uncertainties, the 
existence of a clear-cut definition via the $q$-variance of the 
correlator becomes crucial.

The space-time interpretation of the HBT radius parameters has so far
been largely based on their relations (\ref{2.7}), (\ref{2.10}) with 
the space-time variances of the source which are only true for Gaussian 
correlators. The agreement between the HBT radii from $q$-variances and
from (appropriate) Gaussian fits suggests that these relations continue
to be useful for the space-time interpretation of the correlation 
functions.
%
%
In view of the above agreement between the two types of HBT radius 
parameters, and of our discussion of the interplay between the values 
of $\lambda_i$ and $R_i$ in various types of Gaussian fits to a given 
correlation function, it is not surprising that the intercept 
parameters extracted from Eq.~(\ref{7.7c}) also agree very well with 
the ones extracted from the unidirectional Gaussian fits and shown in 
the right columns of Figs.~\ref{F5} and \ref{F7}. They are therefore not 
presented again.
 
The interesting new information is, of course, contained in the kurtosis
$\Delta_i$ and their $K_\perp$-dependence shown in Fig.~\ref{F12}. In
the side direction the appearance of a non-vanishing (positive) kurtosis 
is clearly linked to the influence of the $\omega$ decays on the 
correlation function and to its visibility in the HBT radius parameter 
$R_s$. This implies that the question whether or not a given 
$K_\perp$-dependence of $R_s$ is caused by resonance decays or not can 
be easily answered by checking the kurtosis of the correlation function. 
If the kurtosis vanishes (or is slightly negative), it is not the 
$\omega$ which causes the $K_\perp$-dependence. At least for the model 
studied here, the kurtosis provides thus the cleanest distinction 
between scenarios with and without transverse flow. Its value and 
$K_\perp$-dependence are thus very important ingredients for the
interpretation of 2-particle correlations.
 
The situation is slightly more complicated in the outward direction:
as long as the source does not expand transversally ($\eta_f=0$),
the visibility of resonance decay effects in $R_o$ is clearly linked 
to a non-zero positive kurtosis of the correlator, and vice versa.
For non-zero transverse flow, however, the outward correlator begins
to develop small deviations from a Gaussian \cite{WSH96} even without
resonance decays; these show up in a {\em negative} value for the 
kurtosis. This effect increases for larger transverse pair momenta 
$K_\perp$.  
 
The kurtosis generated by collective expansion is particularly
prominent in the longitudinal direction where flow-induced non-Gaussian 
features they have been noticed first \cite{WSH96}. The bottom row of 
Fig.~\ref{F12} clearly shows the interplay of non-Gaussian features 
induced by resonance decays (leading to a positive kurtosis) and 
longitudinal expansion flow (causing a negative kurtosis). At small 
$K_\perp$ the resonance contributions dominate; at large $K_\perp$ 
the resonances loose importance while the flow-induced kurtosis 
becomes stronger, leading to overall negative values of the kurtosis. 
 
\section{Conclusions}
\label{sec8}
 
Within a broad class of model emission functions for locally thermalized
and collectively expanding sources we have presented a comprehensive 
study of resonance decay effects on two-pion Bose-Einstein correlations.
We have found that, with regard to their influence on the correlation 
function, the resonances can be subdivided into three classes: 
 \begin{itemize}
 \item
{\em Long-lived resonances} with width $<1$ MeV can not be resolved in
the correlation measurement; they reduce the correlation strength 
$\lambda$ but otherwise do not influence the shape of the correlation 
function in the region where it can be measured. 
 \item
{\em Short-lived resonances} with width $>30$ MeV: they decay into 
pions close to their production point and thus do not change the 
spatial width of the pion emission function. Hence they do not affect
the sideward correlator whose width is defined by the transverse spatial
size of the source. In the outward and longitudinal correlator and in 
the lifetime parameter $R_0$ of the YKP parametrization, which are all 
in one way or other sensitive to the lifetime of the source, they 
contribute via the additional time duration of pion emission due to 
their own lifetime. These contributions are small and on the order of 
the resonance lifetime.
 \item 
{\em The $\omega$ meson.} With its width of about 8 MeV it is not
sufficiently long-lived to escape detection in the correlator, but also
not sufficiently short-lived to not change the spatial width of the
emission function. As a consequence it can lead to severe non-Gaussian
distortions of the correlator.
 \end{itemize}
 
These latter distortions cause serious problems. We have shown in 
Sec.~\ref{sec6} that both the method of extracting width parameters 
from the correlator via Gaussian fits and the calculation of 
these parameters in terms of space-time variances can lead to 
quantitatively unreliable results. The systematic uncertainties 
of Gaussian fits to non-Gaussian correlators were identified
in Sec.~\ref{sec6} (c.f. the discussion following Eq.~(\ref{6.6})) as 
the primary reason for previous claims of much larger resonance effects
on the two-pion HBT radii than found by us. To remove the ambiguities 
associated with non-Gaussian features of the correlator we have 
introduced in Sec.~\ref{sec7} an alternative definition of the HBT 
size parameters and of the intercept parameter $\lambda$ in terms of 
$q$-moments of the correlator which does not rely on the assumption of 
a Gaussian correlator. For sufficiently high statistics data, HBT 
radius parameters determined in this way are free of systematic 
uncertainties. For the examples studied here, they show a much weaker 
influence from resonance decays than we had expected on the basis of
previous work \cite{Marb}. 

The normalized fourth order $q$-cumulant (kurtosis) serves as a 
quantitative lowest-order measure for the non-Gaussian features of 
the correlator. It is sensitive to both resonance decays and flow
which (at least for the models studied here) contribute, however, 
with different signs. The kurtosis thus provides the cleanest signal 
to distinguish between scenarios with and without transverse flow.  
 
Our detailed numerical model study of $q$-moments has shown that 
resonance decays which modify the HBT radius parameters (defined via 
the $q$-variance of the correlator) also lead to a positive 
kurtosis. It can be related to the long non-Gaussian tails in the 
source distribution generated by the decay pions. Collective expansion, 
on the other hand, generates a negative kurtosis because it tends
(in our model) to let the source at its edges decay more steeply than a 
Gaussian. We see practically no flow effects on the kurtosis in
the sideward direction, a weak effect due to transverse expansion in 
the outward direction, and a somewhat larger effect due to the strong 
longitudinal expansion in the longitudinal direction. In the transverse 
direction resonance effects on the HBT radius $R_s$ can thus be directly 
correlated with a non-zero, positive kurtosis. The existence or not of a 
non-vanishing kurtosis $\Delta_s$ and its $K_\perp$-dependence can 
thus be used to assess the amount of contamination in $R_s$ from 
$\omega$-decays and to separate these effects from transverse flow.  
 
$q$-moments thus provide significantly improved information on 
the shape of the correlation function in terms of a still small 
number of relevant parameters $\lambda_i,R_i,\Delta_i$, whose size
and momentum dependence lends itself to an interpretation in terms
of the geometric and dynamic space-time structure of the emitting source.
They are thus expected to further adapt the HBT method to the increased
demand for accuracy in view of the complicated nature of the 
dynamical sources created in relativistic heavy ion collisions and of 
the drastically improved quality of recent correlation measurements. The
new method has been demonstrated to work very well in theory. In view 
of the new high precision data from the Pb-beam at the CERN SPS, it
appears to be experimentally feasible. It will be interesting to see
how far the additional, higher order HBT observables improve our
picture of the spatio-temporal evolution of heavy ion collisions.
 
 
\acknowledgments
 
This work was supported by BMBF, DFG, and GSI. We thank T. Cs\"org\H o, 
P. Foka, M. Ga\'zdzicki, K. Kadija, H. Kalechofsky, M. Martin,
B. Schlei, P. Seyboth, C. Slotta and B. Tom\'a\v{s}ik for many helpful 
conversations. Intensive discussions at the HBT96 Workshop at the
ECT*, Trento, helped to sharpen our arguments; we gratefully 
acknowledge the hospitality of the ECT* and the role it has played in 
crystallizing our thoughts. We would in particular like to acknowledge 
discussions there with S. Voloshin who introduced us to the concept of 
``relative distance distribution'' used in Sec.~\ref{sec7}. One of us 
(U.A.W.) would like to thank S. Kumar, P. Foka and M. Martin for the 
hospitality and help received during visits to Yale and CERN where 
part of this work was written. He also acknowledges a critical 
discussion with M. Gyulassy at Columbia on the use and abuse of
space-time variances. U.H. would like to thank CERN for warm 
hospitality and a stimulating atmosphere during the final stages of 
this work.  
 
\appendix
\section{The emission function for resonance decay pions}
\label{appa}
 
Here, we give details of how to compute the emission function $S_{r\to 
\pi}(x,p)$ for resonance decay pions from a decay channel $r$. We 
follow the treatment in \cite{H63,SKH91} with some notational 
improvements. The resonance $r$ is emitted with momentum $P$ at 
space-time point $X^\mu$ and decays after a proper time $\tau$ at 
$x^\mu = X^\mu + {P^\mu\over M} \tau$ into a pion of momentum $p$ and 
$(n-1)$ other decay products: 
  \begin{equation}
        r \longrightarrow \pi + c_2 + c_3 + ... + c_n \, .
  \label{A1}
  \end{equation}
The decay rate at proper time $\tau$ is 
$\Gamma e^{-\Gamma\tau}$ where $\Gamma$ is the total decay width of 
$r$. Assuming unpolarized resonances with isotropic decay in their rest 
frame, $S_{r\to\pi}(x,p)$ is given in terms of the direct emission function 
$S_r^{\rm dir}(X,P)$ for the resonance $r$ by
 \begin{eqnarray}
   S_{r\to\pi}(x;p) &=& 
        M\, \int_{s_-}^{s_+} ds\, g(s)
        \int{d^3 P \over E_{_P}}\, 
        {\delta}{\left({p\cdot P - E^* M}\right)} 
   \nonumber \\
   &&   \times\, \int d^4X\, \int d\tau \, \Gamma e^{-\Gamma\tau} 
   \nonumber \\
   &&   \times\,
    \delta^{(4)}\left[ x - \left( X + {P\over M} \tau \right) \right]
        S_r^{\rm dir}(X,P)\, .
 \label{A2}
 \end{eqnarray}
Variables with a star denote their values in the 
resonance rest frame, all other variables are given in the fixed 
measurement frame. 
Here $s = \left(\sum_{i=2}^n p_i \right)^2$ is the squared invariant 
mass of the $(n-1)$ unobserved decay products in (\ref{A1}).
It can vary between $s_- = \left( \sum_{i=2}^n m_i \right)^2$ and $s_+ 
= (M-m)^2$. $g(s)$ is the decay phase space for the $(n-1)$ unobserved 
particles. $E^*$ is the energy of the observed decay pion in the 
resonance rest frame and is a function of $s$ only:
 \begin{eqnarray}
   E^* &=& \sqrt{m^2 + {p^*}^2}\, ,
   \nonumber \\
   p^* &=& {\sqrt{[(M+m)^2-s][(M-m)^2-s]} \over 2M}\, .
 \label{A3}
 \end{eqnarray}
We choose for the observer frame a Cartesian coordinate system in 
which the transverse momentum ${\bf p}_\perp$ of the decay pion has 
only an $x$ (``out") and no $y$ (``side") component: 
 \begin{equation}
 \label{A4}
        p^\mu = (E, p_x, p_y, p_{_L}) = 
        \left( m_\perp \cosh y,\, p_\perp, \,
                       0,\, m_\perp \sinh y \right)\, .
 \end{equation}
In this coordinate system the resonance 4-momentum $P$ is parametrized 
by 
 \begin{eqnarray}
         P^\mu &=& (E_{_P}, P_x, P_y, P_{_L}) 
         \nonumber \\
         &=&
        \left( M_\perp \cosh Y,\, P_\perp \cos \Phi, \,
                       P_\perp \sin \Phi,\, M_\perp \sinh Y \right)\, .
 \label{A5}
 \end{eqnarray}
The first $\delta$-function in (\ref{A2}) implements the energy-momentum
constraint $p \cdot P = E^* M$. For $p_{\perp}\ne 0$ it can be used to
fix the azimuthal angle $\Phi$ of the resonance momentum $P$ to
 \begin{eqnarray}
  \Phi_\pm &=& \pm \tilde \Phi \quad \text{with} 
  \nonumber \\
  \cos \tilde \Phi &=& { E\, E_{_P} - p_{_L} P_{_L} - E^* M
                       \over p_\perp P_\perp}
                     \nonumber \\
                   &=& { m_\perp M_\perp \cosh(Y-y) - E^* M
                       \over p_\perp P_\perp}\, .
 \label{A6}
 \end{eqnarray}
We denote by $P^\pm$ the two values of $P$ obtained by inserting 
the two solutions (\ref{A6}) into (\ref{A5}). Rewriting the 
$\delta$-function as $\delta(p \cdot P-E^*M) = \sum_{\pm} 
{\delta(\Phi - \Phi_{\pm}) \over p_\perp P_\perp \sin\Phi_{\pm}}$ and 
doing the $\Phi$-integration in $d^3P/E_{_P} = M_\perp dM_\perp dY\, 
d\Phi$ we find
 \begin{eqnarray}
 \label{A7}
   S_{r\to\pi}(x;p) &=& {1\over 2} \sum_{\pm} \int_{Y_-}^{Y_+} dY
   \int_{M^2_{\perp,-}}^{M^2_{\perp,+}} dM_\perp^2\, \int d^4X 
   \nonumber \\
   && \times
        \int d\tau \, \Gamma e^{-\Gamma\tau} \,
        \delta^{(4)}\left[ x - \left( X + {P^{\pm}\over M} \tau \right) 
        \right]
 \nonumber\\
   && \times\,  S_r^{\rm dir}(X,P^{\pm})\, \Phi_{r\to\pi}(P^{\pm};p)\, ,
 \end{eqnarray}
where 
 \begin{eqnarray}
   && \qquad \Phi_{r\to\pi}(P;p) =
   \nonumber \\
   && \int_{s_-}^{s_+} ds  {M\, g(s) \over 
        \sqrt{P_\perp^2 p_\perp^2 - 
              [E^* M - m_\perp M_\perp \cosh(Y-y)]^2 } }
 \label{A8}
  \end{eqnarray}
is the decay probability for a resonance $r$ with momentum $P$ into
a pion with momentum $p$. It is normalized to the branching ratio 
$b_{r\to\pi}$ for the channel (\ref{A1}) according to 
 \begin{equation}
 \label{A9}
    \int dy \, dp_\perp^2 \, \Phi_{r\to\pi}(P;y,p_\perp) 
    = b_{r\to\pi}  \, .
 \end{equation}
The case $p_\perp = 0$ is a little special: then the constraint 
$p{\cdot}P = E^* M$ in (\ref{A2}) cannot be used to do the 
$\Phi$-integration, but the $M_{\perp}$-integral can be done:
 \begin{eqnarray}
   && S_{r\to\pi}(x;y,p_\perp=0) =
   \nonumber \\
   &&\,\, = M \int_{s_-}^{s_+} ds\, g(s)
        \int_0^{2\pi} d\Phi\,
        \int_{Y_-}^{Y_+} dY\,
        {M\, E^*\over m^2 \cosh^2(Y-y)}
 \nonumber \\
   &&\,\, \times  \int d^4X  \int d\tau \, \Gamma e^{-\Gamma\tau} \,
        \delta^{(4)}\left[ x - \left( X + {P\over M} \tau \right) \right]
         \nonumber \\
   &&\,\, \times 
        S_r^{\rm dir}(X,P)\bigg\vert_{M_\perp = {ME^*\over m\cosh(Y-y)}}\, ,
 \label{A10}
 \end{eqnarray}
In the following we discuss only the case $p_{\perp} \ne 0$. 
The kinematic limits for the integrals in (\ref{A7}) and (\ref{A10})
are, for given $y, m_\perp$ of the decay pion, determined by the zeroes
of the square root in (\ref{A8}): 
  \begin{eqnarray}
        M_{\perp,\pm} &=& \overline{M}_\perp \pm \Delta M_\perp
        \equiv { E^* M m_\perp \cosh(Y-y) \over
            m_\perp^2 \cosh^2(Y-y) - p_\perp^2 } 
          \nonumber \\
         && \pm 
          {M p_\perp 
           \sqrt{ {E^*}^2 + p_\perp^2 - m_\perp^2 \cosh^2(Y-y)}
           \over
           m_\perp^2 \cosh^2(Y-y) - p_\perp^2 }\, ,
    \label{A11} \\ 
        Y_\pm &=& y \pm \Delta Y \equiv 
        y \pm \ln \left( {p^* \over m_\perp} 
                       + \sqrt{ 1 + {{p^*}^2 \over m_\perp^2} } 
                  \right) \, .
    \label{A12}
  \end{eqnarray}
With these ingredients Eq.(\ref{A2}) can be rewritten as
  \begin{eqnarray}
    && S_{r\to\pi}(x,p) 
    = M\, \int_{s_-}^{s_+} ds\, g(s)
        \int_{Y_-}^{Y_+} dY\, 
        \int_{M_{\perp,-}^2}^{M_{\perp,+}^2} dM_\perp^2 
    \nonumber \\
    && \qquad \qquad \times
        \int_0^{\infty} d\tau\, \Gamma e^{-\Gamma\tau} 
  \nonumber \\
    && \times  
       { {1\over 2} \sum_\pm 
         S_r^{\rm dir} \left(x - {P^\pm\over M}\tau, P^\pm \right)
         \over 
         \sqrt{ p_\perp^2 (M_\perp^2 - M^2) -
                [E^* M - m_\perp M_\perp \cosh(Y-y)]^2 } }\, ,
  \label{A13}
 \end{eqnarray}
where the sum is over the two allowed values (\ref{A6}) for $\Phi$.
Rewriting the square root with the help of (\ref{A10}) as
  \begin{equation}
  \label{A14}
        {1\over \sqrt{m_\perp^2 \cosh^2(Y-y) - p_\perp^2}} \,
        {1\over \sqrt{(\Delta M_\perp)^2 - 
                      (M_\perp - \overline{M}_\perp)^2}}
  \end{equation}
and introducing new integration variables $v\in [-1,1]$, 
$\zeta \in [-\pi,\pi]$ via
  \begin{eqnarray}
  \label{A15}
    M_\perp &=& \overline{M}_\perp + \Delta M_\perp \, \cos\zeta\, ,
  \\
  \label{A16}
    Y &=& y  + v\, \Delta Y \, ,
  \end{eqnarray}
Eq.~(\ref{A13}) can be further transformed into
  \begin{equation}
  \label{A17}
        S_{r\to\pi}(x,p) = \sum_\pm \int_{\bf R} 
        \int_0^{\infty}{d\tau}\, \Gamma e^{-\Gamma\tau} 
        S_r^{\rm dir} \left( x -{P^\pm\over M} \tau,P^\pm \right) \, ,
 \end{equation}
with the following shorthand for the integration over the resonance 
momenta:
  \begin{eqnarray}
    \int_{\bf R} &\equiv& M \int_{s_-}^{s_+} ds\, g(s)
        \int_{-1}^1 {\Delta Y \, dv \over 
                   \sqrt{ m_\perp^2 \cosh^2 (v \Delta Y) - p_\perp^2} }
        \nonumber \\
        && \times \int_0^\pi d\zeta 
        \left( \overline{M}_\perp + \Delta M_\perp \cos\zeta \right) \, .
  \label{A18}
  \end{eqnarray}
For the calculation of the correlation function we need the Fourier 
transform of the emission function. It is obtained from (\ref{A17}) 
as
  \begin{eqnarray}
     \tilde S_{r\to\pi}(q,p) 
     &=& \int d^4x\, e^{iq{\cdot}x}\, S_{r\to\pi}(x,p)
     \nonumber \\
        &=& \sum_\pm \int_{\bf R} \int_0^{\infty} d(\Gamma\tau)\, 
        \exp\left[ -\Gamma\tau \left( 1 - i{q{\cdot}P^\pm \over M \Gamma}
                               \right) \right] 
     \nonumber \\
        && \times \int d^4x\, e^{iq{\cdot}x}\, S_r^{\rm dir}(x,P^\pm)
     \nonumber \\
        &=& \sum_\pm \int_{\bf R}
        {1 \over
         1 - i { q{\cdot}P^\pm \over M\Gamma } } \,
        \tilde S_r^{\rm dir}(q,P^\pm) \, ,
  \label{A19}
  \end{eqnarray}
where in the first step we shifted the $x$-integration variable and 
in the second step we performed the $\tau$-integration. For  
two- and three-body decays, this reads
\begin{itemize}
\item
For two-body decays:
  \begin{eqnarray}
  \label{A20}
    g(s) &=& {b\over 4\pi p^*} \delta \left( s - m_2^2 \right)\, .
  \\
   \tilde S_{r\to\pi}(q,p)  &=&  {M b \over 4\pi p^*} \sum_{\pm}
        \int_{-1}^1 {\Delta Y\, dv \over 
             \sqrt{m_\perp^2 \cosh^2(v \Delta Y) - p_\perp^2} }
  \nonumber \\
     && \times
        \int_0^\pi d\zeta {\overline{M}_\perp + \Delta M_\perp \cos\zeta 
             \over 1- i Q_q^\pm }  
        \tilde S_r^{\rm dir}(q,P^\pm)\, ,
  \nonumber \\
   Q^\pm_q &=& {M_\perp\over M\Gamma} 
               \left( q^0 \cosh Y - q_l \sinh Y \right)
  \nonumber \\
            &&  - {P_\perp \over M\Gamma} 
               \left( q_o \cos\Phi_\pm + q_s \sin\Phi_\pm \right) \, .
  \label{A21}
  \end{eqnarray}
 
\item
For three-body decays ($s_- = (m_2+m_3)^2,\, s_+ = (M-m)^2$):
  \begin{equation}
  \label{A22}
        g(s) = {M b\over 2\pi s} 
        {\sqrt{[s - (m_2 + m_3)^2][s - (m_2 - m_3)^2]}
        \over Q(M,m,m_2,m_3)}\, ,
  \end{equation}
  \begin{eqnarray}
        &&\, Q(M,m,m_2,m_3) = \int_{s_-}^{s_+} {ds'\over s'}
        \sqrt{(M+m)^2 - s'}
     \nonumber \\
     && \qquad \times
        \sqrt{s_+ - s'}\sqrt{s_- - s'}
        \sqrt{(m_2-m_3)^2 - s'}\, ,
  \nonumber \\
        &&\tilde S_{r\to\pi}(q,p) 
        = {b M^2 \over 2\pi Q(M,m,m_2,m_3)}
        \nonumber \\
        &&\qquad \times
        \int_{s_-}^{s_+} {ds\over s}\, 
             \sqrt{[s - (m_2 + m_3)^2] [s - (m_2 - m_3)^2]}
  \nonumber \\
        &&\qquad \times \int_{-1}^1  
           {\Delta Y\, dv \over 
            \sqrt{m_\perp^2\cosh^2(v\Delta Y) - p_\perp^2}}
  \nonumber \\
        &&\qquad \times
        \int_0^\pi d\zeta\, {\overline{M}_\perp + \Delta M_\perp \cos\zeta 
             \over 1 - i Q_q^\pm }\, \tilde S_r^{\rm dir}(q,P^\pm)\, .
  \nonumber 
  \end{eqnarray}
\end{itemize}

\section{The Fourier transform of the emission function}
\label{appb}
 
Here, we give details of the calculation of the 
Fourier transform $\tilde S_r^{\rm dir}(q,P) = \int d^4x\, 
e^{iq{\cdot}x} S_r^{\rm dir}(x,P)$ for the resonance emission 
functions (\ref{4.6}). The $\tau$-integration can be done 
analytically: Using $q\cdot x = \tau\, A - q_o x- q_s y$ with $A$ from 
(\ref{5.2c}) we obtain
  \begin{eqnarray}
   && \int \tau\, d\tau\, e^{i A \tau}
        \exp\left(- {(\tau - \tau_0)^2 \over 2(\Delta\tau)^2} \right)
        \nonumber \\
      &&  = \sqrt{2\pi (\Delta\tau)^2} \, e^{i A \tau_0} \,
        e^{- {1\over 2}A^2 (\Delta\tau)^2}\,
        \left(\tau_0 + iA(\Delta\tau)^2\right)\, .
  \label{B1}
  \end{eqnarray}
The angular integral is also easily done: writing $q_o= q_\perp 
\cos\varphi$, $q_s = q_\perp \sin\varphi$, such that $q_o x + q_s y =
r q_\perp \cos (\phi-\varphi)$ (where $\phi$ is the polar angle of $x$ 
and $y$), the integral over the angle-dependent part of the source 
function (\ref{4.6}) is written as
 \begin{eqnarray}
   &&\int_0^{2\pi} d\phi\, e^{- i r q_\perp \cos(\phi-\varphi)}\,
        e^{{P_\perp\over T} \sinh\eta_t \cos(\phi-\Phi)}
        \nonumber \\
   && = \int_0^{2\pi} d\psi\, e^{- i r q_\perp \cos(\psi+\tilde\varphi)}\,
        e^{{P_\perp\over T} \sinh\eta_t \cos\psi}\, ,
 \label{B2}
 \end{eqnarray}
with $\psi=\phi-\Phi$, $\tilde \varphi = \Phi - \varphi$. Separating 
real and imaginary parts one obtains modified Bessel functions 
\cite{AS65}:
  \begin{mathletters}
  \label{B3} 
  \begin{eqnarray}
    &&\int_0^{2\pi} d\psi\, 
    \exp\left( {\textstyle{P_\perp \over T}}\sinh\eta_t\cos\psi \right)
    \nonumber \\
    &&\qquad \times\, \cos\left( r q_\perp\cos\tilde\varphi\cos\psi
                  - r q_\perp\sin\tilde\varphi\sin\psi \right)
  \nonumber \\
    && = \pi\, \left( I_0(\sqrt{C-iD}) + I_0(\sqrt{C+iD}) \right)\, ,
  \label{B3a} \\
    && -i \int_0^{2\pi} d\psi\,
    \exp\left( {\textstyle{P_\perp \over T}}\sinh\eta_t\cos\psi \right)
    \nonumber \\
    &&\qquad \times\, \sin\left( r q_\perp\cos\tilde\varphi\cos\psi
                  - r q_\perp\sin\tilde\varphi\sin\psi \right)
  \nonumber \\
    && = \pi\, \left( I_0(\sqrt{C-iD}) - I_0(\sqrt{C+iD}) \right)\, ,
  \label{B3b}
  \end{eqnarray}
  \end{mathletters}
where $C$ and $D$ are given in (\ref{5.2}a,b). The remaining integrals 
over $r$ and $\eta$ are given in (\ref{5.1}) and must be done 
numerically.
 
The single particle spectrum is obtained by evaluating $\tilde S(q,P)$ 
at $q=0$. Then also $A$, $A_q$ and $D$ vanish (i.e. the dependence on 
the polar angle $\Phi$ of the transverse momentum ${\bf P}_\perp$ 
drops out), and $C$ reduces to $C = (P_\perp \sinh\eta_t(r)/T)^2$. The 
transverse momentum spectrum is obtained by additionally integrating 
over the rapidity $Y$ associated with $P$. This integral can again be 
done analytically:
  \begin{eqnarray}
  \label{B4}
    {dN_r^{\rm dir}\over dM_\perp^2} 
    &=& \pi \int dY\, \tilde S_r^{\rm dir}(0;M_\perp,Y) 
    \nonumber \\
    &=&
        {2J_r+1 \over (2\pi)^{3/2}} \, 
        M_\perp \tau_0\, e^{{\mu_r\over T}} \,
        \int_0^\infty r\, dr \, e^{-{r^2\over 2R^2}}\,
  \nonumber\\
    && \times 
        I_0\left( {\textstyle{P_\perp\over T}}\sinh\eta_t(r) \right)
        \int d\eta\,\exp\left(-{\eta^2\over 
                                      2(\Delta\eta)^2} \right) 
  \nonumber\\
    && \times
       \int dY\, \cosh(\eta-Y)
  \nonumber\\
    && \times 
       \exp\left[ -{\textstyle{M_\perp\over T}}
                   \cosh\eta_t\cosh(\eta-Y) \right]
  \nonumber \\
   &=& {2J_r+1 \over 2\pi} \, 
        (2 \tau_0 \Delta\eta)\,
        e^{{\mu_i\over T}}\,M_\perp 
        \int_0^\infty r\, dr \, e^{-{r^2\over 2R^2}}\,
        \nonumber \\
   && \times  K_1\left( {\textstyle{M_\perp\over T}}\cosh\eta_t(r) \right)\,
        I_0\left( {\textstyle{P_\perp\over T}}\sinh\eta_t(r) \right) \, .
  \label{B4a}
  \end{eqnarray}
The $K_1$-function results from the last integral in (\ref{B4})
after a simple shift of the integration variable, and the remaining 
Gaussian integral over $\eta$ is trivial.


\end{document}